\documentclass[a4paper,11pt]{article}
\pdfoutput=1 

\usepackage{jheppub} 

\usepackage[T1]{fontenc} 
\usepackage[textwidth=1.5cm, size=small]{todonotes}
\usepackage{slashed}
\usepackage{graphics,bm}
\usepackage{epsfig}
\usepackage{graphicx}
\usepackage{braket}
\usepackage{cancel} 
\usepackage{amsmath}
\usepackage{amssymb}
\usepackage{bbold}
\usepackage{wrapfig}
\usepackage{mathtools}
\usepackage{color}
\newcommand{\beq}{\begin{equation}}
\newcommand{\eeq}{\end{equation}}
\newcommand{\bqa}{\begin{eqnarray}}
\newcommand{\eqa}{\end{eqnarray}}

\def\square{\vcenter{\vbox{\hrule height.4pt
          \hbox{\vrule width.4pt height4pt
          \kern4pt\vrule width.3pt}\hrule height.4pt}}}

\newcommand*\diff{\mathop{}\!\mathrm{d}}
\DeclareMathOperator{\Tr}{tr}

\title{Quantum corrections to slow-roll inflation: scalar and tensor modes
}

\author[a]{Jens O. Andersen,}
\author[a,b,1]{Magdalena Eriksson,\note{Corresponding author.}}
\author[b]{Anders Tranberg}

\affiliation[a]{Department of Physics, Faculty of Natural Sciences, NTNU, 
Norwegian University of Science and Technology, H{\o}gskoleringen 5,
N-7491 Trondheim, Norway}
\affiliation[b]{Faculty of Science and Technology, University of Stavanger,
  4036 Stavanger, Norway}

\emailAdd{andersen@tf.phys.ntnu.no}
\emailAdd{magdalena.eriksson@ntnu.no}
\emailAdd{anders.tranberg@uis.no}

\abstract{Inflation is often described through the dynamics of a scalar field, slow-rolling in a suitable potential. Ultimately, this inflaton must be identified with the expectation value of a quantum field, evolving in a quantum effective potential. The shape of this potential is determined by the underlying tree-level potential, dressed by quantum corrections from the scalar field itself and the metric perturbations. Following \cite{Herranen:2015aja}, we compute the effective scalar field equations and the corrected Friedmann equations to quadratic order in both scalar field, scalar metric and tensor perturbations. We identify the quantum corrections from different sources at leading order in slow-roll, and estimate their magnitude in benchmark models of inflation. We comment on the implications of non-minimal coupling to gravity in this context.
}
\begin{document} 
\maketitle

\section{Introduction}\label{sec:introduction}

Cosmological observations show that the very early Universe underwent a stage of accelerating expansion \cite{Aghanim:2018eyx}. This is possible when the matter in the Universe exhibits some specific thermodynamical properties, often phrased in terms of the equation of state. An inflationary equation of state emerges naturally, if during this inflationary epoch, the thermodynamics was dominated by a scalar field degree of freedom (fundamental or composite), evolving in an appropriate potential. Fairly generically, if the field is initially displaced far from its equilibrium value, it will slow-roll back to the potential minimum in such a way that inflation is achieved.
In this well-known and elegant formalism, the homogeneous scalar field is treated as a single classical degree of freedom $\phi$, and the combined system of field equation and Einstein (Friedmann) equations for the cosmological expansion may be readily solved (see e.g. \cite{Lyth:2009zz}). In a homogeneous 
Friedmann-Lema\^{i}tre-Robertson-Walker (FLRW) background, field and metric perturbations are introduced, and their spectra can be computed in a straightforward way. Through a well-established procedure, these primordial quantum fluctuations can be shown to seed the temperature fluctuations of the CMB, and hence the formation of structure in the Universe. Direct comparison with observations allows us to constrain the parameters and form of the scalar field potential \cite{Akrami:2018odb}.

Ultimately, the scalar field must be treated quantum mechanically, and then an adjustment of the terminology is required. The object $\phi$ may be identified as the time-dependent expectation value of the scalar field.\footnote{Or, according to taste, as the mean field, order parameter, one-point function or condensate.} This degree of freedom now evolves in the quantum effective potential, and the minimum of this potential corresponds to the thermodynamic equilibrium state. 

An important distinction is that this is not an effective potential in the sense of a low-energy effective theory, where degrees of freedom above a certain cutoff have been integrated out. Such effective potentials are often invoked in inflationary model-building to motivate a wide range of functional forms of classical potentials. The quantum effective potential is the free energy of the system, once all quantum and thermal fluctuations have been included at all scales (see for instance \cite{peskin}), and for cosmological applications it includes gravitational corrections as well. From the quantum effective action, equations of motion for both the scalar mean field and the metric may be derived, and these then include all quantum corrections and hence all the thermodynamical information for the system.\footnote{The quantum effective action
reduces to the quantum effective potential if the mean fields are constant, in which case it is given by the classical
potential plus the quantum corrections to it.} 

In practice, computing this quantum effective potential in an expanding cosmological background is possible, but less straightforward than in Minkowski space (see e.g. \cite{Birrell,toms}). As in Minkowski space, an expansion in terms of Feynman diagrams is introduced and truncated. The issue of renormalisation arises, and just as for Minkowski space, consistency limits the set of viable tree-level potentials. 

An alternative, but equivalent approach to the quantum dynamics of inflation is to introduce quantum corrections at the level of the operator equations of motion. Starting from the classical Friedmann and scalar field equations, we may expand in powers of the perturbations around the FLRW solution, and by taking quantum averages generate a Schwinger-Dyson-like set of evolution equations for the correlators. These may then in principle be solved (see for instance \cite{Markkanen:2012rh,Herranen:2016xsy}). 

As a consequence of gravity's highly non-linear properties, including metric perturbations in this scheme is technically challenging, and wholesale resummations have not been performed. The most common avenue is to include the effect of the homogeneous time-dependent metric in the computation of scalar field correlators and quantum effective actions, leading to higher order curvature contributions to the system. One recurring result from this procedure is that the ''potential'' appearing in the energy, the pressure and the force of the Friedmann and scalar field equations receive different corrections. This means that the ubiquitous slow-roll formalism must be adjusted, and a number of standard identities correct to some order in a slow-roll expansion, are in addition also correct only at tree-level (or some order in perturbations) \cite{Herranen:2013raa,Bilandzic:2007nb}. 

In \cite{Herranen:2015aja}, scalar metric perturbations were introduced in a computation of quantum-corrected Friedmann and scalar field equations, to leading order in slow-roll and leading order in fluctuations. The constrained quantisation of the field and metric degrees of freedom was given particular emphasis, and the result was a manageable set of evolution equations. The scalar fluctuations came out slow-roll suppressed, and the numerical effect of including them turned out to be negligible, except near the end of inflation, and then only for small-field inflation. 

In the present work, we revisit this picture and include the tensor perturbations, for which we compute the corresponding leading order corrections to both scalar field evolution and Friedmann equations. We will find that these new corrections enter differently in the evolution than the scalar metric perturbations, and that they are larger than the scalar corrections.

We also consider non-minimal coupling between the scalar field and the curvature, and although we will not carry this complication along throughout the calculation, we will be able to illustrate how a non-minimal coupling may affect the results.

Quantum effects during and after inflation have received significant attention in many different contexts and in many different guises. Prominent examples are non-Gaussian effects in cosmological observables due to self-interactions of metric perturbations (see e.g. \cite{Maldacena:2002vr,Bartolo:2004if}); logarithmic infrared divergences of correlators in de Sitter and slow-roll backgrounds and methods to handle them (see \cite{Bartolo:2007ti,Seery:2010kh} for a review), including the stochastic approximation \cite{Starobinsky:1994bd,Tsamis:2005hd}; and effective actions for the inflaton field itself (without metric perturbations), in various resummation schemes, both in an adiabatic expansion (around Minkowski space $H=0$) \cite{toms,Tranberg:2008ae} and a slow-roll expansion (around de Sitter space $\dot{H}=0$) \cite{Sloth:2006az,Seery:2007we,Riotto:2008mv,Herranen:2013raa}, as well as in strict de Sitter space (see e.g. \cite{Serreau:2011fu,Gautier:2013aoa,Gautier:2015pca}).

What we are concerned with here is yet another context; quantum corrections to evolution equations of the inflaton and metric between horizon exit and the end of inflation, taking into account both field and metric fluctuations. These fluctuations influence the relation between the slow-roll parameters at the end of inflation and at horizon crossing, which in turn enter in observables in the sky. 

The paper is organised as follows. In section \ref{sec:backgroundeqs} we set up our action, the standard Friedmann and scalar equations at tree-level. We then introduce metric fluctuations in Newtonian gauge and compute the field equations for them. These are then quantised in sections \ref{sec:quanttensor} and \ref{sec:quantscalar}. We proceed to derive the corrected Friedmann and scalar field equation to second order in perturbations in section \ref{sec:quadraticOrder} and discuss renormalisation. In section \ref{sec:magnitude}, we briefly consider the magnitude of corrections and estimate the effect of quantum corrections on the inflationary dynamics. We conclude in section \ref{sec:conclusion}.
Details of the calculations are listed in Appendices \ref{Appendix:CommutationRelations}--\ref{App:CalcCorrs}
for completeness. 

\section{Background and field equations in Newtonian gauge}
\label{sec:backgroundeqs}

We consider a single inflaton field $\phi$ non-minimally coupled to gravity with action of the form (following the sign convention (+++) of \cite{Wheeler}) 
\begin{equation}
    S = \int\diff^4x\sqrt{-g}\left[\tfrac{1}{2}M_\text{Pl}^2F({\phi})R-\tfrac{1}{2}{g}^{\mu\nu}{\phi}_{,\mu}{\phi}_{,\nu}-V({\phi})\right]\ , \label{TheAction}
\end{equation}
where $M_\text{Pl}^2\equiv (8\pi G)^{-1}$ is the reduced Planck mass and where for the moment $F(\phi)$ and the tree-level potential $V(\phi)$ are kept general.\footnote{ In some cases, it can be convenient by a conformal transformation to perform the calculation in the Einstein frame and then transform the results back again. In the Einstein frame the non-minimal coupling is absent but the field potential and normalisation are different. We have chosen to stay in the Jordan frame throughout. } 
The scalar field $\phi$ lives in a spacetime defined by the metric $\hat{g}_{\mu\nu}$, and we will assume that field and metric may be written as
\begin{equation}
\begin{aligned}
    \phi(x) &= \phi(t)+\delta\phi(t,x^i)\ , \\
    \hat{g}_{\mu\nu}(x) &= g_{\mu\nu}(t)+\delta g_{\mu\nu}(t,x^i)\ ,
    \label{eq:fluct}
\end{aligned}
\end{equation}
where $g_{\mu\nu}$ is the flat FLRW metric; $g_{\mu\nu}=\textrm{diag}[-1,a^2(t)\delta_{ij}]$, with $a(t)$ being the scale factor and where $t$ denotes cosmic time. Latin indices run from one to three and Greek indices from zero to three.

The metric perturbations may be decomposed into scalar, vector, and tensor modes. It is common to immediately discard the vector perturbations, as they decay away in an expanding Universe. We will do so here as well. Choosing to work in Newtonian gauge, we may write for the line element
\begin{equation}
    \diff s^2 = -(1+2\Phi)\diff t^2+a^2[1+2(-\Psi\delta_{ij}+E_{ij})]\diff x^i\diff x^j\ , \label{eq:pert}
\end{equation}
where $\Phi$ and $\Psi$ are scalar potentials and $E_{ij}$ is a traceless, transverse matrix containing the tensor perturbations, i.e. 
$E_{ij}$ satisfies
$\partial^iE_{ij}=0$ and $E^i{}_i=0$.

The Einstein field equations are obtained by variation of the action in Eq. \eqref{TheAction} with respect to the metric, which yields
\begin{equation}
    G_{\mu\nu} \equiv R_{\mu\nu}-\tfrac{1}{2}Rg_{\mu\nu}=\frac{1}{M_\text{Pl}^2}\frac{1}{F(\phi)}\Tilde{T}_{\mu\nu}\equiv \frac{1}{M_\text{Pl}^2}{T}_{\mu\nu}\ , 
    \label{eq:EFEgeneral}
\end{equation}
where 
\begin{equation}
    \tilde{T}_{\mu\nu} \equiv \phi_{,\mu}\phi_{,\nu}-g_{\mu\nu}[\tfrac{1}{2}g^{\rho\kappa}\phi_{,\rho}\phi_{,\kappa}+V(\phi)]+ F(\phi)_{;\mu,\nu}-g_{\mu\nu}\Box F(\phi)\ .
    \label{eq:EMT}
\end{equation}
Here and in the following a comma subscript denotes a partial derivative, a semi-colon denotes a covariant derivative and the ''box'' operator is the d'Alembertian in curved spacetime
\begin{equation}
\Box = \frac{1}{\sqrt{-g}}\partial_\mu\sqrt{-g}\partial^\mu\ .    
\end{equation}

The effective energy-momentum tensor ${T}_{\mu\nu}$ is covariantly conserved, and we will assume that $F(\phi)\neq 0$.
Variation of the action with respect to $\phi$ yields the scalar field equation
 \begin{equation}
     \Box{\phi}+\tfrac{1}{2}M_\text{Pl}^2F_{,{\phi}}R-V_{,{\phi}}=0\ .\label{eq:EOMInflatonGeneral}
 \end{equation}
Next, we insert the perturbed fields of Eqs. (\ref{eq:fluct}) and (\ref{eq:pert}) into Eqs. (\ref{eq:EFEgeneral})--(\ref{eq:EOMInflatonGeneral}) and extract equations to zeroth order in perturbations (the ''classical'' Friedmann and inflaton equations); to first order in perturbations (the mode equations for the field and metric scalar and tensor perturbations in the background FLRW metric); and to quadratic order in perturbations (the ''quantum-corrected'' Friedmann and inflaton equations).
 
The zeroth-order equations will provide a slow-roll background in which to solve the first-order (linear) mode equations. These will in turn be quantised and inserted into the quadratic-order equations, in order to explicitly compute the quantum corrections to the cosmological evolution. 

\subsection{Zeroth order in perturbations: the classical equations}
\label{sec:0thorder}

To zeroth order the Friedmann equation \eqref{eq:EFEgeneral} and the scalar field equation \eqref{eq:EOMInflatonGeneral} reduce to the familiar expressions
\begin{align}
    3H^2 &= \frac{1}{M_\text{pl}^2}\frac{1}{F}\left(\tfrac{1}{2}\dot{{\phi}}^2+V\right)-3H\frac{\dot{F}}{F}\ , \label{BackgroundFriedmannEnergyDensity}\\ 
    2\Dot{H}+3H^2 &= -\frac{1}{M_\text{Pl}^2}\frac{1}{F}\left(\tfrac{1}{2}\dot{{\phi}}^2-V\right)-2H\frac{\dot{F}}{F}-\frac{\Ddot{F}}{F}\ , \label{BackgroundFriedmannPressure}\\
    0 &= \ddot{\phi}+3H\dot{\phi}+V_{,\phi}-3M_\text{Pl}^2F_{,\phi}(\Dot{H}+2H^2) \ , \label{TreeLevelEoM} 
\end{align}
where $H=\dot{a}/a$ and the right-hand side of Eqs. (\ref{BackgroundFriedmannEnergyDensity}) and (\ref{BackgroundFriedmannPressure}) represents energy density and pressure, respectively. Here the dots are derivatives with respect to cosmic time 
$t$ and we have inserted $R=6(\Dot{H}+2H^2)$, the FLRW Ricci scalar. 
We note that in the limit of Einstein gravity, $F(\phi)=1$, we recover the standard Friedmann equations. Also, since we assume that $F\equiv F(\phi)$, we have $\dot{F}\propto \dot{\phi}$, $\ddot{F}\propto \ddot{\phi},\dot{\phi}^2$. We also note that the same object $V$ (or $V_{,\phi}$) appears in all three equations, which then allows for the direct application of the slow-roll formalism. 

It will therefore be convenient to introduce the 
dimensionless slow-roll parameters
\begin{align}
    \epsilon_H=-\frac{\dot{H}}{H^2}=1-\frac{\mathcal{H}'}{\mathcal{H}^2}\ ,\qquad \delta_H=-\frac{\ddot{\phi}}{H\dot{\phi}}=1-\frac{{\phi}''}{\mathcal{H}{\phi}'}\ ,\qquad \epsilon_F=\frac{1}{2}\frac{\dot{F}}{HF}=\frac{1}{2}\frac{F'}{\mathcal{H}F}\ .\label{SRparams}
\end{align}
The primes refer to derivatives with respect to conformal time $\eta$, where
\begin{align}
    a(\eta)\diff \eta = \diff t\ ,\qquad \eta =\int\frac{\diff t}{a}\ ,\qquad \mathcal{H}=\frac{a'}{a}\ .
\end{align}
Inflation is equivalent to the slow-roll parameters being small, i.e. less than unity, and evolving slowly in time. 

One may apply the slow-roll formalism to solve Eqs. (\ref{BackgroundFriedmannEnergyDensity})--(\ref{TreeLevelEoM}) to any order in slow-roll parameters one chooses. However, as was emphasised in \cite{Herranen:2013raa,Bilandzic:2007nb}, at the quantum level, corrections manifest themselves differently in different relations, and the effective potential $V$ is no longer the same. Hence, when including quantum corrections,
some of the elegance of the slow-roll formalism is lost. We will see this explicitly below.

\subsection{First order in perturbations: equations of motion for the fluctuations} \label{sec:firstorder}

The perturbations of the scalar field and the metric Eq. (\ref{eq:fluct}) give
rise to linear perturbations in $F$, $G_{\mu\nu}$ and $\tilde{T}_{\mu\nu}$, which we denote by
$\delta F$, $\delta G_{\mu\nu}$ and $\delta \tilde{T}_{\mu\nu}$, respectively. 
The Einstein equation can then be written as 
\begin{align}
(F+\delta F)(G_{\mu\nu}+\delta G_{\mu\nu})=\frac{1}{M_\text{Pl}^2}(\tilde{T}_{\mu\nu}+\delta\tilde{T}_{\mu\nu}), \label{eq:firstOrderPertEinsteinEq}
\end{align}
Using the zeroth-order equations, one immediately finds (now in mixed index notation)
\begin{equation}
    \delta G^\mu{}_\nu = \frac{1}{M_\text{Pl}^2}\frac{1}{F}\left(\delta\tilde{T}^\mu{}_\nu-\frac{\delta F}{F}\tilde{T}^\mu{}_\nu\right)\ , \label{eq:firstOrderPertEinsteinEqMinimal}
\end{equation}
with the understanding, that since $F$ is a function of $\phi$ only, $\delta F\equiv F_{,\phi}\delta\phi$. The 00- and $0i$-components of Eq. \eqref{eq:firstOrderPertEinsteinEqMinimal} are
\bqa\nonumber
    3H(\dot{\Psi}+H\Phi) -\frac{1}{a^2}\nabla^2\Psi &=& \frac{1}{2M_\text{Pl}^2F}\left( -\dot{\phi}\dot{\delta\phi}+\dot{\phi}^2\Phi-V_{,\phi}\delta\phi\right)
    -\frac{3\dot{F}}{2F}(\dot{\Psi}+2H\Phi)
    \\&&
    \label{LinearEnergyDensity}
    +\frac{3H}{2}\frac{\dot{\delta F}}{F}-\frac{1}{2a^2}\nabla^2\frac{\delta F}{F}+\frac{1}{2}\frac{\delta F}{F}\left( \frac{\tfrac{1}{2}\dot{{\phi}}^2+V}{M_\text{Pl}^2F}-3H\frac{\dot{F}}{F}\right)\;,\nonumber\\
    \\
    \dot{\Psi}+H\Phi &=& \frac{1}{2M_\text{Pl}^2F}\dot{\phi}\delta\phi+\frac{1}{2}\frac{\dot{\delta F}}{F}-\frac{H}{2}\frac{\delta F}{F}-\frac{1}{2}\frac{\dot{F}}{F}\Phi\ . 
    \label{LinearMixed}
\eqa
We note that in the minimal coupling case $F(\phi)=1$, the relations simplify substantially.
The $ii$-component gives us
\bqa\nonumber
    \ddot{\Psi}+3H\dot{\Psi}+H\dot{\Phi}+(2\dot{H}+3H^2)\Phi
    &=& \frac{1}{2M_\text{Pl}^2F}\Big(\dot{\phi}\dot{\delta\phi}-\dot{\phi}^2\Phi-V_{,\phi}\delta\phi\Big)
      -\frac{\ddot{F}}{F}\Phi+H\frac{\dot{\delta F}}{F}
      \\ && \nonumber
      -\frac{1}{2}\frac{\dot{F}}{F}(4H\Phi+\dot{\Phi}+2\dot{\Psi})
      \nonumber
      +\frac{1}{2}\frac{\ddot{\delta F}}{F}
      -\frac{1}{2a^2}\nabla^2\left(\frac{\delta F}{F}\right)
      \\&&
      -\frac{1}{2}\frac{\delta F}{F}\left( \frac{\tfrac{1}{2}\dot{\phi}^2-V}{M^2_\text{Pl}F}+2H\frac{\dot{F}}{F}+\frac{\ddot{F}}{F}\right)\ , \label{eq:perturbedPressure}
  \eqa
  Again, the non-minimal coupling is responsible for substantial complexity. For the off-diagonal components $i\neq j$ of the perturbed Einstein equation (\ref{eq:firstOrderPertEinsteinEqMinimal}), we may treat the longitudinal and transverse components separately, to find
\begin{align}
    \frac{1}{a^2}\partial^i\partial_j(\Psi-\Phi) &= \partial^i\partial_j\left(\frac{1}{a^2}\frac{\delta F}{F}\right)\ , \label{LinearScalarAnisotropicStress} \\
    \Ddot{E}^i{}_j+3H\dot{E}^i{}_j-\frac{1}{a^2}\nabla^2 E^i{}_j &= -\frac{\dot{F}}{F}\dot{E}^i{}_j\ . \label{TensorEFE}
\end{align}
We make a few important observations. The right-hand side of Eq. (\ref{LinearScalarAnisotropicStress}) is an anisotropic stress, and in the presence of a space-dependent non-minimal coupling, it is nonzero. The equality $\Psi=\Phi$ is often used to simplify the system of equations, but since $\delta F\propto \delta\phi$ is in general space-dependent, this is no longer possible. In the absence of non-minimal coupling ($F(\phi)=1$), this relation is recovered. 
For Einstein gravity, Eq. (\ref{TensorEFE}) is a free wave equation with cosmological redshift. However, once $F$ becomes time-dependent, an additional damping/amplification is introduced, with explicit dependence on $\dot{\phi}$. The new damping is proportional to the slow-roll parameter $\epsilon_F$.

By combining Eqs. (\ref{LinearEnergyDensity}) and
(\ref{eq:perturbedPressure}), we arrive at the scalar mode equation 
\begin{align}
    \ddot{\Psi}+6H\dot{\Psi}+&H\dot{\Phi}+2(\dot{H}+3H^2)\Phi-\frac{1}{a^2}\nabla^2\Psi = -\frac{V_{,\phi}\delta\phi}{M_\text{Pl}^2F}-\frac{\ddot{F}}{F}\Phi-\frac{\dot{F}}{2F}\left[5(2H\Phi+\dot{\Psi})+\dot{\Phi}\right]\nonumber\\
    &+\frac{1}{2}\frac{\ddot{\delta F}}{F}+\frac{5H}{2}\frac{\dot{\delta F}}{F}-\frac{1}{a^2}\nabla^2\left(\frac{\delta F}{F}\right)\label{eq:GeneralScalarModeEq}
    +\frac{1}{2}\frac{\delta F}{F}\left(\frac{2V}{M_\text{Pl}^2F}-\frac{\ddot{F}}{F}-5H\frac{\dot{F}}{F}\right)\ , 
\end{align}
and inserting Eq. \eqref{LinearScalarAnisotropicStress} into Eq. \eqref{LinearMixed}, we obtain 
\begin{equation}
    \Dot{\Psi}+\left(H+\frac{\Dot{F}}{2F}\right)\Psi = \frac{1}{2}\left( \frac{1}{M_\text{Pl}^2F}\Dot{\phi}\delta\phi+\frac{\Dot{\delta F}}{F}+H\frac{\delta F}{F}+\frac{\Dot{F}\delta F}{F^2}\right)\ .\label{PsiDeltaPhiRelation}
\end{equation}
At this stage, one procedure could be to solve Eq. (\ref{PsiDeltaPhiRelation}) for $\delta\phi$ and substitute the result 
into Eq. (\ref{eq:GeneralScalarModeEq}). 
An elegant field redefinition allows us to absorb the anisotropic stress in Eq. \eqref{LinearScalarAnisotropicStress} into new variables
 $\check{\Psi}$ and $\check{\Phi}$,
\begin{equation}
    \check{\Psi}\equiv  -\Psi+\frac{\delta F}{2F}\ , \qquad \check{\Phi}\equiv\Phi+\frac{\delta F}{2F}\ , \qquad \check{\Psi}+\check{\Phi}=0\ , \label{eq:FancyVariables}
\end{equation}
so that Eq. \eqref{eq:GeneralScalarModeEq} becomes homogeneous \cite{Hwang:1990re,hwang0}. This is what we need to carry out the quantisation of the modes. However, eventually, we will be interested in the Friedmann and inflaton equations to quadratic order, and these turn out not to be easily written in terms of these new variables (see Appendix \ref{Appendix:Feqs}). In effect, the procedure is ruined in the general case by the appearance of $\dot{\delta\phi}$ in Eq. (\ref{PsiDeltaPhiRelation}).

The equation of motion for the scalar field perturbations is found to be 
\begin{equation}
    \ddot{\delta\phi}+3H\dot{\delta\phi}-\frac{1}{a^2}\nabla^2\delta\phi+V_{,\phi\phi}\delta\phi-\tfrac{1}{2}M_\text{Pl}^2RF_{,\phi\phi}\delta\phi = -2V_{,\phi}\Phi+\dot{\phi}(\dot{\Phi}+3\dot{\Psi})+M_\text{Pl}^2F_{,\phi}(R\Phi+\tfrac{1}{2}\delta R)\ , \label{DeltaphiEoM}
\end{equation}
where the perturbation of the Ricci scalar is given by
\begin{equation}
     \delta R = -12(\dot{H}+2H^2)\Phi-6H(\dot{\Phi}+4\dot{\Psi})-6\ddot{\Psi}+\frac{2}{a^2}\nabla^2(2\Psi-\Phi)\ . \label{deltaR}
\end{equation}
The above equations capture the dynamics of the fluctuating fields, 
and we will proceed by quantising them in the next two sections.

\subsection{Quantisation of the tensor modes}
\label{sec:quanttensor}

The field equations for the inflaton fluctuations and the metric scalar perturbations are too complicated to quantise in the case of a general $F$. But the tensor fluctuation equation is relatively peaceful, so we will consider that first, before we specialise to $F(\phi)=1$.
We have from Eq. (\ref{TensorEFE}) that in conformal time
\begin{align}
E_{ij}''+2\mathcal{H}\left(1+\epsilon_F\right)E_{ij}'-\nabla^2E_{ij}=0\ .\label{eq:TensorFieldEoM}
\end{align}
Introducing $z_g=\sqrt{F}a$ and $v\delta_{ij}=z_gE_{ij}$ (for each instance of $i,j$), this reduces in momentum space $v(q)$ to
\begin{align}
\label{eq:tensormode}
v''+\left(q^2-\frac{z_g''}{z_g}\right)v=0\ .
\end{align}
Using the slow-roll expression for conformal time
\begin{align}
\eta = -\frac{1}{aH}\frac{1}{1-\epsilon_H}=-\frac{1}{\mathcal{H}}\frac{1}{1-\epsilon_H}\ ,
\end{align}
we find
\begin{align}
\frac{z_g''}{z_g}=\frac{1}{\eta^2}\frac{(1+\epsilon_F)(2-\epsilon_H+\epsilon_F)}{(1-\epsilon_H)^2}\equiv \frac{n}{\eta^2}\ .
\end{align}
Treating $n$ as a constant, the solution to Eq. (\ref{eq:tensormode}) is
\begin{align}
    v(\eta,q)=\sqrt{|\eta|}\left[c_1({\bf q})H_\nu^{(1)}(q|\eta|)+c_2({\bf q})H_\nu^{(2)}(q|\eta|)\right] \ , \qquad \nu = \sqrt{n+\frac{1}{4}}\ ,
\end{align}
where $H_{\nu}^{(1)}$ and $H_{\nu}^{(2)}$ are Hankel functions of the
first and second kind, respectively.
Choosing the Bunch-Davies vacuum, the coefficients are set to $c_1({\bf q})=0$ and $c_2({\bf q})=1$ \cite{Birrell}.
We still need the overall normalisation of the mode and so we write the quantised tensor field as
\begin{align}
E_{ij}(\eta,{\bf x})=\int\frac{\diff^3q}{(2\pi)^{3/2}}\sum_{\lambda=+,\times}\left[\hat{a}_{\lambda{\bf q}}\tilde{h}_{\lambda{\bf q}}(\eta)e_{ij}^{(\lambda)}({\bf q})e^{i{\bf q\cdot x}}+\textrm{h.c.}\right]\ ,  \label{eq:Eansatz}
\end{align}
where  ${a}_{\lambda{\bf q}}$ and ${a}^{\dagger}_{\lambda{\bf q}}$
are annihilation and creation operators and $\tilde{h}_{\lambda{\bf q}}$  are scalar functions. The subscripts 
$+,\times$ refer to the two polarisation states of the polarisation tensors $e^{(\lambda)}_{ij}$, that satisfy
\begin{align}
e^{(\lambda)}_{ij}({\bf q})e^{(\lambda')}_{ij}({\bf q})=2\delta_{\lambda\lambda'}\ , \qquad e^{(\lambda)}_{ij}=e_{ji}^{(\lambda)}\ ,\qquad e^{(\lambda)i}{}_i=0\ , \qquad \partial^ie_{ij}^{(\lambda)}=0\ .
\end{align}
The creation and annihilation operators satisfy the commutation relation
\begin{align}
\label{eq:acommutator}
[\hat{a}_{\lambda{\bf q}},\hat{a}^\dagger_{\lambda{\bf q'}}]=\delta_{\lambda\lambda'}\delta^3({\bf q-q'})\ ,
\end{align}
with all other commutators vanishing. 
Inserting the ansatz of Eq. (\ref{eq:Eansatz}) into the action, one finds that the quadratic contributions $S_\text{t}$ from the tensor may be written 
\begin{align}
S_\text{t} = M_\text{Pl}^2\int \diff\eta \diff^3x a^2 F\sum_{\lambda}[\hat{h}^{'2}-\hat{h}_{\lambda}{}^{,i}\hat{h}_{\lambda,i}]\ ,
\end{align}
where 
\begin{align}
\hat{h}_\lambda(\eta,{\bf x})=   \int\frac{\diff^3q}{(2\pi)^{3/2}}\left[\hat{a}_{\lambda{\bf q}}\tilde{h}_{\lambda{\bf q}}(\eta)e^{i{\bf q\cdot x}}+\textrm{h.c.}\right]  \ .
\end{align}
The canonical quantisation condition is 
\begin{align}
[\hat{h}_{\lambda}(\eta,{\bf}),\hat{\pi}_\lambda(\eta,{\bf x'})]=i\delta^3({\bf x-x'})\ ,
\end{align}
and defining the conjugate momentum
\begin{align}
\hat{\pi}_{\lambda}(\eta,{\bf x})=\frac{\partial \mathcal{L}}{\partial \hat{h}_{\lambda}^{'}}=2a^2 M_\text{Pl}^2F\hat{h}_{\lambda}^{'}\ ,   
\end{align}
together with Eq. (\ref{eq:acommutator}), we may write down the Wronskian for the mode functions themselves as
\begin{align}
\tilde{h}_{\lambda{\bf q}}(\eta)\tilde{h}^{'*}_{\lambda{\bf q}}(\eta)-  \tilde{h}^*_{\lambda{\bf q}}(\eta)\tilde{h}^{'}_{\lambda{\bf q}}(\eta)  =\frac{i}{2a^2 M_\text{Pl}^2F}\ .
\end{align}
This gives us the final normalisation 
\begin{align}
    \tilde{h}_{\lambda{\bf q}}(\eta) = \sqrt{\frac{\pi |\eta|}{8a^2 M_\text{Pl}^2F}}H_{\nu}^{(2)}(q|\eta|)\ , \label{tensorModeFunction}
\end{align}
where $\nu$ is approximated as
\begin{equation}
    \nu \approx \tfrac{3}{2}+\epsilon_F+\epsilon_H+\mathcal{O}(\epsilon^2)\ ,
\end{equation}
to first order in slow-roll parameters. We see that the effect of non-minimal coupling on the evolution of the tensor modes enters through $\epsilon_F$ and the overall normalisation.

\subsection{Quantisation of the scalar modes}
\label{sec:quantscalar}

Since it is not possible to express all scalar-perturbation dependence in the quadratic-order quantum corrections solely in terms of the new variable $\hat{\Psi}$ of Eq. \eqref{eq:FancyVariables}, we now consider the Einstein gravity limit where $F(\phi)=1$. In this case, Eq. \eqref{eq:GeneralScalarModeEq} reduces to
\begin{equation}
    \Ddot{\Psi}+H(1+2\delta_H)\Dot{\Psi}+2H^2(\delta_H-\epsilon_H)\Psi-\frac{1}{a^2}\nabla^2\Psi = 0\ , \label{eq:MetricScalarEoM}
\end{equation}
using Eqs. (\ref{TreeLevelEoM}) and
(\ref{PsiDeltaPhiRelation}) and the slow-roll parameters in Eq. \eqref{SRparams}.
The scalar field equations (\ref{LinearEnergyDensity})--(\ref{LinearScalarAnisotropicStress}) in combination with the equation of motion for $\delta\phi$ in Eq. \eqref{DeltaphiEoM} is an overdetermined system, and hence requires a constrained quantisation procedure \cite{Dirac:1950pj}. 
To this end we introduce the canonical momenta
\begin{align}
    \pi_\Phi &\equiv \frac{\partial\mathcal{L}}{\partial\Phi'} = 0\ , \\
    \pi_\Psi &\equiv \frac{\partial\mathcal{L}}{\partial\Psi'} = -6a^2M_\text{Pl}^2[\Psi'+\mathcal{H}(\Phi+\Psi)]\ , \label{phiPsi}\\
    \pi_{\delta\phi} &\equiv \frac{\partial\mathcal{L}}{\partial\delta\phi'} = a^2[\delta\phi'-\phi'(\Phi+3\Psi)]\ , \label{phiDeltaphi}
\end{align}
and encode the constraints of the field equations by introducing the conjugate momenta 
\begin{align}
    \chi_1 &\equiv \pi_\Phi\ , \\
    \chi_2 &\equiv \nabla^2\Psi+3\mathcal{H}'\Psi-\frac{1}{2M_\text{Pl}^2}\left(a^2V_{,\phi}\delta\phi+\frac{\phi'}{a^2}\pi_{\delta\phi}-\frac{\mathcal{H}}{a^2}\pi_\Psi \right) \ , \\
    \chi_3 &\equiv \mathcal{H}\Psi+\frac{1}{2M_\text{Pl}^2}\left(\phi'\delta\phi+\frac{1}{3a^2}\pi_\Psi\right)\ , \\
    \chi_4 &\equiv \Phi-\Psi\ , 
\end{align}
where then
\begin{equation}
    \chi_\alpha=0\ , \qquad \alpha=1,2,3,4.
\end{equation}
The constrained quantisation is realised by way of the Dirac bracket 
defined as
\begin{equation}
    [A,B]_\text{D} \equiv [A,B]_\text{P}-[A,\chi_m]_\text{P}(C^{-1})_{mn}[\chi_n,B]_{\text{P}}\ , \qquad m,n=2,3,
\end{equation}
where the Poisson bracket is defined as
\begin{equation}
    [A,B]_\text{P}\equiv \sum_{\varphi=\Psi,\delta\phi}\left(\frac{\partial A}{\partial \varphi}\frac{\partial B}{\partial\Pi_\varphi}-\frac{\partial B}{\partial\varphi}\frac{\partial A}{\partial \Pi_\varphi}\right) \ , 
\end{equation}
and $C_{mn}$ is a non-singular constraint matrix 
\begin{equation}
    C_{mn}\equiv [\chi_m,\chi_n]_\text{P}\ . 
\end{equation}
The quantised variables will then have equal-time commutation relations given by
\begin{equation}
    [A,B] = i[A,B]_\text{D}\ ,
\end{equation}
for which we obtain
\begin{equation}
     [\Psi(\mathbf{x}),\Psi(\mathbf{x'})] = [\delta\phi(\mathbf{x}),\delta\phi(\mathbf{x'})] = 0\ , \qquad [\Psi(\mathbf{x}),\Psi'(\mathbf{x'})] = i\frac{\phi'^2}{4M_\text{Pl}^4a^2\nabla^2}\delta^3(\mathbf{x}-\mathbf{x'})\ ,\label{eq:MainCorrelationRelationScalars}
\end{equation}
and where the resulting commutation relations for $\Psi,\,\delta\phi$ combined with the conjugate momenta $\pi_\Psi,\,\pi_{\delta\phi}$ are listed in Appendix \ref{Appendix:CommutationRelations}.  

The field $\Psi$ can be promoted to an operator and decomposed in terms of mode functions $\tilde{f}_{\mathbf{k}}(\eta)$ as
\begin{equation}
    \hat{\Psi}(\eta,\mathbf{x}) = \int\frac{\diff^3 k}{(2\pi)^{3/2}}\left(\hat{b}_{\mathbf{k}}\tilde{f}_\mathbf{k}(\eta)e^{i\mathbf{k}\cdot\mathbf{x}}+\text{h.c.}\right) \ , 
\end{equation}
where the creation and annihilation operators $\hat{b}_{\mathbf{k}},\hat{b}_{\mathbf{k}}^\dagger$ fulfil the standard commutation relations 
\begin{equation}
    [\hat{b}_{\mathbf{k}},\hat{b}_{\mathbf{k'}}]=[\hat{b}_{\mathbf{k}}^\dagger,\hat{b}_{\mathbf{k'}}^\dagger]= 0 \ ,\qquad [\hat{b}_{\mathbf{k}},\hat{b}_{\mathbf{k'}}^\dagger]=\delta^3(\mathbf{k}-\mathbf{k'}) \ ,\label{eq:MainScalarCommutators}
\end{equation}
so that the equation of motion for the mode functions becomes  
\begin{equation}
    \tilde{f}_\mathbf{k}''-\frac{2\delta_H}{\eta}\tilde{f}_\mathbf{k}'+\left(\frac{2(\delta_H-\epsilon_H)}{\eta^2}+k^2\right)\tilde{f}_\mathbf{k} = 0\ . \label{ScalarModeFunctionEoM}
\end{equation}
Here we have approximated the Hubble parameter $\mathcal{H}\simeq-1/\eta$ to zeroth order in slow-roll. From the equal-time commutation relation in Eq. \eqref{eq:MainScalarCommutators}, the Wronskian is calculated to
\begin{equation}
    \tilde{f}_\mathbf{k}(\eta)\tilde{f}_\mathbf{k}'^*(\eta)-\tilde{f}_\mathbf{k}^*(\eta)\tilde{f}_\mathbf{k}'(\eta) = i\left(\frac{\phi'}{2M_\text{Pl}^2a|\mathbf{k}|}\right)^2\ .\label{eq:scalarWronskian}
\end{equation}
The solution to the mode equation \eqref{ScalarModeFunctionEoM} is then given by 
\begin{equation}
    \tilde{f}_\mathbf{q}(\eta) = \frac{\sqrt{\pi|\eta|}}{2}\frac{\phi'}{2M_\text{Pl}^2a|\mathbf{k}|}H_{\upsilon}^{(2)}(k|\eta|)=\sqrt{\frac{\pi\epsilon_H}{8}}\frac{\mathcal{H}}{M_\text{Pl}^2a|\mathbf{k}|}(-\eta)^{1/2}H_{\upsilon}^{(2)}(-k\eta)\ , \label{scalarModeFunction} 
\end{equation}
where the index $\upsilon$ is approximated as
\begin{equation}
    \upsilon = \tfrac{1}{2}\sqrt{1+8\epsilon_H-4\delta_H}\approx \tfrac{1}{2}+2\epsilon_H-\delta_H\ .\label{upsilondefinition}
\end{equation}

\section{Quantum-corrected equations of motion in the Einstein-gravity limit} \label{sec:quadraticOrder}

Taking the vacuum expectation value of the perturbed equations of motion, we calculate the quantum corrections in the Einstein gravity limit $F(\phi)=1$. The second-order terms of the Friedmann and mean-field equations are listed in Appendix \ref{Appendix:Feqs} (with $F$ general), where the number of terms reduces significantly after taking the vacuum expectation value. In the end, the equations may be expressed in terms of just a few two-point correlators, which we will therefore consider first.

\subsection{Correlation functions and renormalisation}
\label{sec:QuantumCorrected:CorrelatorFunctionsRenormalisation}

The quantum-corrected Friedmann and mean-field equations can be expressed in terms of the two-point correlation functions $\braket{\varphi^2},\,\braket{\varphi\nabla^2\varphi}$ and $\braket{\varphi\nabla^4\varphi}$, with $\varphi=E_{ij},\Psi$, by using the relations in Appendix \ref{App:CorrRels}. Having obtained the mode solutions for the tensors in Eq. \eqref{tensorModeFunction} and scalars in Eq. \eqref{scalarModeFunction}, these correlators can now be calculated explicitly. By use of the field decompositions in Eqs. (\ref{eq:Eansatz}) and
(\ref{eq:acommutator}), we obtain for the tensor correlators:
\begin{align}
  \ \, &\begin{aligned}
    \mathllap{C_0} &\equiv\braket{E_{ij}E_{ij}}= \frac{1}{2\pi^2}\frac{H^2}{M_\text{Pl}^2F}\bigg[\left( \frac{1}{2(\epsilon_H+\epsilon_F)}+\log2+\psi(\tfrac{3}{2})\right)\left(-1+\Lambda_\text{IR}^{-2(\epsilon_H+\epsilon_F)} \right)\\
      &\hspace{4.5cm}+\tfrac{1}{2}\Lambda_\text{UV}^2+\log\Lambda_\text{UV}\bigg]+\mathcal{O}(\epsilon)
    \ , \label{C0correlator}
  \end{aligned}\\
    \mathllap{C_2} &\equiv \braket{E_{ij}\nabla^2E_{ij}} =- \frac{1}{4\pi^2}\frac{a^2H^4}{M_\text{Pl}^2F}\left[\tfrac{1}{2}\Lambda_{\text{UV}}^4+\Lambda_{\text{UV}}^2-\tfrac{1}{2}\Lambda_\text{IR}^4-\Lambda_\text{IR}^2\right]\ ,
\label{c2}
\end{align}
where $\psi(x)$ denotes the digamma function and $\Lambda_{\rm IR,UV}$ are infrared and ultraviolet cutoffs respectively, see the details in Appendix \ref{App:CalcCorrs}. For the tensor modes, we have for illustration  retained the non-minimal coupling through $\epsilon_F$ and the overall normalisation $1/F$.

For the scalars we obtain
\begin{align}
    &D_0 \equiv \braket{\Psi^2}=\frac{\epsilon_HH^2}{ 8\pi^2M_\text{Pl}^2}\left[\left( \frac{1}{2(2\epsilon_H-\delta_H)}+\log2+\psi(\tfrac{1}{2})\right)\left(-1+\Lambda_\text{IR}^{-2(2\epsilon_H-\delta_H)}\right)+\log\Lambda_\text{UV}\right]+\mathcal{O}(\epsilon)\ , \label{D0correlator} \\
    \label{d2}
    &D_2 \equiv \braket{\Psi\nabla^2\Psi}= -\frac{\epsilon_Ha^2{H}^4}{ 16\pi^2M_\text{Pl}^2}\left(\Lambda_\text{UV}^2-\Lambda_\text{IR}^2\right)+\mathcal{O}(\epsilon)\ , \\
    &D_4 \equiv \braket{\Psi\nabla^4\Psi}= \frac{\epsilon_Ha^4{H}^6}{ 32\pi^2M_\text{Pl}^2}\left(\Lambda_\text{UV}^4-\Lambda_\text{IR}^4\right)+\mathcal{O}(\epsilon)\ .
    \label{d4}
\end{align}

We notice that the scalar correlators each have an overall factor of $\epsilon_H$, which is absent in the tensor correlators. This is a direct consequence of the mode function normalisation, where the constrained quantisation procedure of the scalars gives an additional factor of $\phi'^2$ in the commutation relations in Eq. \eqref{eq:MainCorrelationRelationScalars}.
However, we will see that once the correlators are reinstated in the equations of motion, they appear at the same order in slow-roll.

The correlators have both IR and UV divergences. A careful treatment of the UV involves applying dimensional regularisation, which removes all power law divergences, and where logarithmic divergences are absorbed into counterterms for higher-order invariant operators $R^2,R^{\mu\nu}R_{\mu\nu}$, etc. \cite{Markkanen:2013nwa}. The computations in Appendix \ref{App:CalcCorrs} employ a simpler cutoff regularisation to identify where UV and IR divergences appear, after which we assume that the UV can be dealt with, so that only the physical IR effects remain. Based on this discussion, 
the tensor correlators will hereafter be taken to be
\begin{equation}
\begin{aligned}
    C_0 &= -\frac{H^2}{4\pi^2 M_\text{Pl}^2F(\epsilon_H+\epsilon_F)}\left(1-\Lambda_\text{IR}^{-2(\epsilon_H+\epsilon_F)} \right)\ , \\
    C_2 &= \frac{a^2H^4}{4\pi^2 M_\text{Pl}^2F}(\tfrac{1}{2}\Lambda_\text{IR}^4+\Lambda_\text{IR}^2)\ , \label{eq:RenormalisedTensorCorrs}
\end{aligned}
\end{equation}
and the scalar correlators by
\begin{equation}
\begin{aligned}
    D_0 &= -\frac{\epsilon_HH^2}{ 16\pi^2M_\text{Pl}^2(2\epsilon_H-\delta_H)}\left(1-\Lambda_\text{IR}^{-2(2\epsilon_H-\delta_H)}\right)\ , \\
    D_2 &= \frac{\epsilon_Ha^2{H}^4}{ 16\pi^2M_\text{Pl}^2}\Lambda_\text{IR}^2\ , \\ 
    D_4 &= -\frac{\epsilon_Ha^4{H}^6}{ 32\pi^2M_\text{Pl}^2}\Lambda_\text{IR}^4\ . \label{eq:RenormalisedScalarCorrs}
\end{aligned}
\end{equation}
Furthermore, we should only keep terms of leading order in slow-roll, since the mode functions are only valid to that order.  The correlators $C_2,D_2$ and $D_4$ are proportional to the infrared cutoff, and as we imagine $\Lambda_\text{IR}\ll 1$, they give negligible contributions (and are set to zero). This leaves the IR-divergent $C_0,D_0$, for which we may write
\begin{equation}
\begin{aligned}
1-\Lambda_\text{IR}^{-2(\epsilon_H+\epsilon_F)}
&=2(\epsilon_H+\epsilon_F)|\log\Lambda_\text{IR}|+\mathcal{O}(\epsilon^2)\ ,
\\
    1-\Lambda_\text{IR}^{-2(2\epsilon_H-\delta_H)}&= 2(2\epsilon_H-\delta_H)|\log\Lambda_\text{IR}|+\mathcal{O}(\epsilon^2)\ .
\end{aligned}
\end{equation}
This yields
\bqa
C_0=-\frac{1}{2\pi^2}\frac{H^2}{M_\text{Pl}^2F}|\log\Lambda_\text{IR}| +\mathcal{O}(\epsilon)\ ,\\
    D_0=-\frac{\epsilon_H}{8\pi^2}\frac{H^2}{ M_\text{Pl}^2}|\log\Lambda_\text{IR}| +\mathcal{O}(\epsilon^2)\ .
\eqa
We will in the following count $C_0$ as $\mathcal{O}(1)$ and $D_0$ as $\mathcal{O}(\epsilon)$. Below we will have a further analysis of the magnitude of the IR-cutoff.

\subsection{Friedmann equations}\label{sec:friedmanneq}

Perturbing the Einstein equations to quadratic order in fluctuations, the quantum-corrected equation will take the form
\begin{equation}
    G^\mu{}_\nu = \frac{1}{M_\text{Pl}^2}(\tilde{T}^\mu{}_\nu+\braket{\delta_2\tilde{T}^\mu{}_\nu})-\braket{\delta_2G^\mu{}_\nu}\ , \label{eq:secondOrderPertEinsteinEq}
\end{equation}
where the quadratic-order contributions $\delta_2\tilde{T}_{\mu\nu}$ and $\delta_2G_{\mu\nu}$ are given in Appendix \ref{Appendix:Feqs}. For the 00-component of Eq. \eqref{eq:secondOrderPertEinsteinEq}, we obtain
\begin{equation}
    \begin{aligned}
    3H^2 &= \frac{1}{M_\text{Pl}^2}\left[\tfrac{1}{2}\dot{\sigma}^2+V-\braket{\delta_2\tilde{T}^0{}_0}\right] +\braket{\delta_2G^0{}_0}\\
    &= \frac{1}{M_\text{Pl}^2}\left[\tfrac{1}{2}\dot{\sigma}^2+V+\tfrac{1}{2}\braket{\dot{\delta\phi}^2}+\frac{1}{2a^2}\braket{(\nabla\delta\phi)^2}-2\dot{\phi}\braket{\Phi\dot{\delta\phi}}+2\dot{\phi}^2\braket{\Phi^2}+\tfrac{1}{2}V_{,\phi\phi}\braket{\delta\phi^2}\right]\\
    &\quad+\tfrac{1}{2}\braket{\dot{E}^{ij}\dot{E}_{ij}}+4H\braket{E^{ij}\dot{E}_{ij}}-12H^2D_0-3\braket{\dot{\Psi}^2}-\frac{1}{a^2}\left[5D_2+\tfrac{3}{2}C_2\right]\\
    &=  \frac{1}{M_\text{Pl}^2}\left(\tfrac{1}{2}\dot{\sigma}^2+V\right) + \frac{H^2}{\epsilon_H}\bigg[\tfrac{1}{4}\epsilon_H\frac{\partial_t^2C_0}{H^2}+\tfrac{19}{8}\epsilon_H\frac{\partial_tC_0}{H}-\tfrac{7}{4}\epsilon_H\frac{1}{a^2}\frac{C_2}{H^2}\\
    &\quad+(-\tfrac{3}{2}\epsilon_H+\tfrac{1}{2}\delta_M)\frac{\partial_t^2D_0}{H^2}+(-6\delta_H+\tfrac{9}{2}\epsilon_H+\tfrac{3}{2}\delta_M)\frac{\partial_tD_0}{H}+(-12\epsilon_H+\delta_M)D_0\\
    &\quad-\frac{1}{a^2}\frac{\partial_t^2D_2}{H^4}-(2+\delta_H)\frac{1}{a^2}\frac{\partial_tD_2}{H^3}-(1+2\delta_H+4\epsilon_H+\delta_M)\frac{1}{a^2}\frac{D_2}{H^2}+\frac{1}{a^4}D_4\bigg]\ ,
\end{aligned}\label{FinalFriedmannTime}
\end{equation}
where we used the constraint relations in Eqs. (\ref{LinearMixed}) and
(\ref{LinearScalarAnisotropicStress}) to express $\Phi$ and $\delta\phi$ in terms of $\Psi$. Furthermore, we have introduced a new slow-roll parameter
\begin{equation}
    \delta_M \equiv  \frac{V_{,\phi\phi}}{H^2} \simeq 3(\delta_H+\epsilon_H)\ , \label{eq:DeltaMSRparameter}
\end{equation}
where the last relation follows from Eqs. (\ref{TreeLevelEoM}) and (\ref{SRparams}), and is correct at leading order in slow-roll. We have also used that 
\begin{equation}
    \epsilon_H H^2 = \frac{\dot{\phi}^2}{2M_\text{Pl}^2}\ ,\label{phiDotInTermsOfEpsilonH}
\end{equation}
which can be obtained by combining Eqs. (\ref{BackgroundFriedmannEnergyDensity}) and (\ref{BackgroundFriedmannPressure}) with Eq. \eqref{SRparams}. For the spatial component of the Friedmann equations with $i=j$, we get 
\begin{equation}
    \begin{aligned}
    2\dot{H}+3H^2 
     &=\frac{1}{M_\text{Pl}^2}\left[-\tfrac{1}{2}\dot{\phi}^2+V-\tfrac{1}{2}\braket{\dot{\delta\phi}^2}-\frac{1}{2a^2}\braket{(\nabla\delta\phi)^2}+2\dot{\phi}\braket{\Phi\dot{\delta\phi}}-2\dot{\phi}^2\braket{\Phi^2}+\tfrac{1}{2}V_{,\phi\phi}\braket{\delta\phi^2}\right]\\
     &\quad-\tfrac{1}{2} \braket{\dot{E}^{ij} \dot{E}_{ij}} -\frac{5}{2a^2}\braket{E^{ij}\nabla^2E_{ij}}  -4(2\dot{H}+3H^2)\braket{\Psi}^2 -8H \braket{\Psi \dot{\Psi}}-\braket{\dot{\Psi}^2}+\frac{1}{a^2}\braket{\Psi\nabla^2\Psi}\\
     &=\frac{1}{M_\text{Pl}^2}\left(-\tfrac{1}{2}\dot{\phi}^2+V\right)+ \frac{H^2}{\epsilon_H}\bigg[ -\tfrac{1}{4}\epsilon_H\frac{\partial_t^2C_0}{H^2}-\tfrac{3}{8}\epsilon_H\frac{\partial_tC_0}{H}-\tfrac{9}{4}\epsilon_H\frac{1}{a^2}\frac{C_2}{H^2}\\
     &\quad+(-\tfrac{1}{2}\epsilon_H+\delta_M)\frac{\partial_t^2D_0}{H^2}+(6\delta_H-\tfrac{21}{2}\epsilon_H+\tfrac{3}{2}\delta_M)\frac{\partial_tD_0}{H}+(-12\epsilon_H+\delta_M)D_0\\
     &\quad+\frac{1}{a^2}\frac{\partial_t^2D_2}{H^4}+(2+\delta_H)\frac{1}{a^2}\frac{\partial_tD_2}{H^3}+(1+2\delta_H+4\epsilon_H-\delta_M)\frac{1}{a^2}\frac{D_2}{H^2}-\frac{1}{a^4}\frac{D_4}{H^4}\bigg]\ . 
\end{aligned}\label{FinalFriedmannSpace}
\end{equation}
For illustration, we have included contributions to leading order in slow-roll and quantum corrections, but also (parts of) the higher order contributions in slow-roll. However, for consistency, we must again truncate the whole expression at leading order. Remembering also that $D_2\simeq D_4\simeq C_2\simeq 0$, we then find:
\bqa
    3H^2 
    &=&  \frac{1}{M_\text{Pl}^2}\left(\tfrac{1}{2}\dot{\phi}^2+V\right) + H^2\left(-12+\frac{\delta_M}{\epsilon_H}\right)D_0+\tfrac{19}{8}H\partial_tC_0\ ,
\label{FinalFriedmannTimeLead}
\\
    2\dot{H}+3H^2 
     &=&\frac{1}{M_\text{Pl}^2}\left(-\tfrac{1}{2}\dot{\phi}^2+V\right)+ H^2\left(-12+\frac{\delta_M}{\epsilon_H}\right)D_0 -\tfrac{3}{8}H\partial_tC_0\ .
\label{FinalFriedmannSpaceLead}
\eqa
We recall that $C_0\sim\mathcal{O}(1)$, while $D_0\sim\mathcal{O}(\epsilon)\sim\partial_t C_0$. 
Hence, we see that scalar and tensor contributions to the effective potential appear at the same order $\mathcal{O}(\epsilon)$, even though the correlators themselves are of different order. 

The corrections to the classical relations involve terms that could equally be grouped with the left-hand side ($\propto H^2$) and the right-hand side ($\phi$-dependent). Interpreting the quantum corrections as corrections to the potential, we see that the resulting effective potential is different in the two Friedmann equations. At this level in slow-roll, this difference originates from the tensor contributions only, while the scalar contributions are the same.

\subsection{Mean-field equation}\label{sec:meanfield}

The perturbed mean-field equation \eqref{eq:EOMInflatonGeneral} is displayed in Eq. \eqref{MeanFieldEqBackgroundAndQuadraticTermsNonMinimal}, which in the Einstein gravity limit reduces to
\bqa\nonumber
0 &=&  \ddot{\phi}+3H\dot{\phi}+V_{,\phi} + \ddot{\phi}(6\Psi^2+E^i{}_jE^j{}_i)+\dot{\phi}\big[ 3H(6\Psi^2+E^i{}_jE^j{}_i)+12\Psi\dot{\Psi}+2\dot{E}^i{}_jE^j{}_i\big]\\
\nonumber
        && -4\ddot{\delta\phi}\Psi-4\dot{\delta\phi}[3H\Psi+\dot{\Psi}]+\frac{2}{a^2}\partial_i[E^{ij}\partial_j\delta\phi] -V_{,\phi}(2\Psi^2+E^i{}_jE^j{}_i)
        \\&& 
        -2V_{,\phi\phi}\Psi\delta\phi+\tfrac{1}{2}V_{,\phi\phi\phi}\delta\phi^2 \; .
\eqa
Taking the vacuum expectation value of this, we arrive at the quantum-corrected equation
\begin{equation}
\begin{aligned}
    0 &= \Ddot{\phi}\left[1+\frac{1}{\epsilon_H}\left(\epsilon_HC_0+(14\epsilon_H-8\delta_H)D_0-4(1+\delta_H)\frac{\partial_t D_0}{H}-2\frac{\partial_t^2D_0}{H^2}+\frac{4}{a^2}\frac{D_2}{H^2}\right) \right]\\
    &\quad+3H\Dot{\phi}\bigg[1+\frac{1}{3\epsilon_H}\left(3\epsilon_HC_0+\epsilon_H\partial_tC_0+(18\epsilon_H+2\delta_M)D_0-(4\epsilon_H-\delta_M)\frac{\partial_tD_0}{H}-\frac{4}{a^2}\frac{D_2}{H^2}-\frac{4}{a^2}\frac{\partial_tD_2}{H^3}\right) \bigg]\\
    &\quad+V_{,\phi}(1+6D_0-C_0)+\frac{M_\text{Pl}^{2}}{\epsilon_H}V_{,\phi\phi\phi}\left[(1+2\delta_H-2\epsilon_H)D_0+(\tfrac{3}{2}+\delta_H)\frac{\partial_t D_0}{H}+\frac{1}{2}\frac{\partial_t^2D_0}{H^2}-\frac{1}{a^2}\frac{D_2}{H
   ^2}\right]\ . \label{MeanFieldEqBackgroundAndQuadraticTerms}
\end{aligned}
\end{equation}
This equation can be derived via two routes. Either using the field equation \eqref{LinearMixed} to re-express $\ddot{\delta\phi}$ in terms of $\Psi$, or by using Eq. \eqref{DeltaphiEoM}. Here we have chosen the latter way, as it is more straightforward, although both methods must yield the same result at leading order in slow-roll parameters. 
We may further truncate in slow-roll, and again set $D_2\simeq D_4\simeq C_2\simeq 0$, to obtain
\begin{equation}
\begin{aligned}
    0 &= \Ddot{\phi}\left[1+C_0+\left(14-8\frac{\delta_H}{\epsilon_H}\right)D_0\right]
    +3H\Dot{\phi}\bigg[1+C_0+\tfrac{1}{3}\partial_tC_0+\left(6\epsilon_H+\frac{2}{3}\frac{\delta_M}{\epsilon_H}\right)D_0 \bigg]\\
    &\quad+V_{,\phi}(1-C_0+6D_0)+M_\text{Pl}^{2}V_{,\phi\phi\phi}\frac{1+2\delta_H-2\epsilon_H}{\epsilon_H}D_0\ . 
\end{aligned}
\end{equation}
Here we notice that the quantum corrections already appear at $\mathcal{O}(1)$, and treating $\ddot{\phi}$ as higher order, we may further truncate to get
\begin{equation}
\begin{aligned}
    0 &= 3H\Dot{\phi}\left(1+C_0\right)
    +V_{,\phi}(1-C_0)+M_\text{Pl}^{2}V_{,\phi\phi\phi}\frac{D_0}{\epsilon_H}\ . \label{MeanFieldEqLeading}
\end{aligned}
\end{equation}
As for the Friedmann equation, corrections from scalars and tensor enter at the same order, but the tensor contributions now enter as a multiplicative correction to the potential and the damping (Hubble) rate.

Comparing to the classical field equation (where $C_0=D_0=0$), it is not possible to identify the corrections as simply modifying the potential. On the other hand, we note that for $C_0=0$ (ignoring tensors), when comparing to Eq. (\ref{FinalFriedmannTimeLead}), the term $V_{,\phi\phi\phi}D_0/\epsilon$ may be related to (the derivative of) $\delta_M D_0/\epsilon$, provided $D_0/\epsilon$ is assumed to not depend on $\phi$. Once $C_0\neq 0$ this correspondence is lost, showing that tensor modes make a qualitative difference.

\section{Magnitude of corrections: examples}
\label{sec:magnitude}

In this section, we estimate the magnitude of the quantum corrections to the mean-field and Friedmann equations for large-field monomial inflation and quartic hilltop inflation. We wish to compute their values during inflation, and we choose the time of horizon crossing, which we take to be $N=50-60$ $e$-foldings before the end of inflation.

First, we need a prescription to evaluate the IR-divergent correlators. As noted in section \ref{sec:QuantumCorrected:CorrelatorFunctionsRenormalisation}, with a small IR cutoff $\Lambda_\text{IR}\rightarrow 0$, all correlators become negligible except $C_0$ and $D_0$, which diverge logarithmically. The IR logarithm can be related to the number of $e$-foldings during inflation. If we assume that the IR cutoff is set to exclude superhorizon modes from the loop integrals, i.e. so that the comoving momenta is cut off by the initial Hubble radius as $k\geq a_\text{in}H_\text{in}$, then for $x\equiv -k\eta\geq \Lambda_\text{IR}$ the cutoff is 
\begin{equation}
    \Lambda_\text{IR} = \frac{a_\text{in}H_\text{in}}{aH(1-\epsilon_H)} \ . \label{eq:LambdaIRGeneral}
\end{equation}
With approximately constant $\epsilon_H$, the scale factor and Hubble parameter are solved by
\begin{equation}
    a=a_\text{in}e^N\ , \qquad H=H_\text{in}e^{-\epsilon_HN}\ , \qquad N\equiv \int_{t_\text{in}}^t\diff t'H(t') \ ,\label{aHsolutions}
\end{equation}
where $N$ is the number of $e$-foldings from horizon exit to the end of inflation. The assumption $\epsilon_H\simeq \text{const.}$ signifies that we are working at leading order in slow-roll. Inserting the solutions of Eq. \eqref{aHsolutions} into Eq. \eqref{eq:LambdaIRGeneral}, the IR limit can be written
\begin{equation}
    \Lambda_\text{IR} = \frac{1}{1-\epsilon_H}e^{-(1-\epsilon_H)N}\ , 
\end{equation}
so that for $\Lambda_\text{IR}\ll 1$ its logarithm can be approximated to 
\begin{equation}
    |\log\Lambda_\text{IR}| \simeq (1-\epsilon_H)N\ . \label{eq:logLambdaIR}
\end{equation}
The role of IR-divergences and ways to deal with them are discussed for instance in \cite{Tsamis:2005hd,Seery:2010kh}.

Now, when comparing the magnitude of the tensor and scalar corrections,  we must consider the minimal-coupling limit in which the tensor correlator of Eq. \eqref{eq:RenormalisedTensorCorrs} is given by
\begin{equation}
    C_0 =  \frac{1}{2\pi^2}\frac{H^2}{M_\text{Pl}^2}N_\text{t}\ ,
    \label{eq:tensorCorrMagGeneral}
\end{equation}
where we have introduced a number $N_\text{t}$ related to $N$ as 
\begin{equation}
    N_\text{t}\equiv -\frac{1}{2\epsilon_H}\left(1-e^{2\epsilon_H|\log\Lambda_\text{IR}|}\right)\simeq-\frac{1}{2\epsilon_H}\left(1-e^{2\epsilon_H N}\right)\ .\label{eq:Nt}
\end{equation}
Similarly for the scalars, we write 
\begin{equation}
    D_0 = \frac{\epsilon_H}{8\pi^2}\frac{H^2}{ M_\text{Pl}^2}N_\text{s} \ , \label{eq:scalarCorrMagGeneral}
\end{equation}
introducing the quantity $N_\text{s}$ according to
\begin{equation}
N_\text{s}\equiv -\frac{1}{2(2\epsilon_H-\delta_H)}\left(1-e^{2(2\epsilon_H-\delta_H)|\log\Lambda_\text{IR}|}\right)\simeq-\frac{1}{2(2\epsilon_H-\delta_H)}\left(1-e^{2(2\epsilon_H-\delta_H)N}\right)\ .\label{eq:Ns}
\end{equation}
We can relate the combination of slow-roll parameters to the measured spectral index (see below for the definition of the ''potential'' slow-roll parameters $\epsilon_V$, $\delta_V$) according to
\begin{equation}
    -0.035=n_\text{s}-1\equiv \tfrac{2}{3}\delta_M-6\epsilon_H=2\delta_H-4\epsilon_H\simeq 2\delta_V-6\epsilon_V\ . \label{eq:spectralIndexSRCombination}
\end{equation}
Using this in Eq. \eqref{eq:Ns}, we find for the range $N=50-60$ that $N_\text{s}\simeq 136-205$. It is less straightforward to infer $\epsilon_H$ from $n_\text{s}$. If $\delta_H\simeq 2\epsilon_H\simeq (n_\text{s}-1)/2$, then $N_\text{t}=33-37$ or $80-108$ depending on the sign of $\epsilon_H$. There are also models where $\epsilon_H\ll\delta_H$ during inflation, in which case we may expand $N_\text{t}$ so that $N_\text{t}\simeq N$. We conclude that $N_\text{s,t}\simeq100$, within a factor of a few. This will be sufficient for our estimates.

\subsection{Monomial models}\label{sec:monomial}

The simplicity of monomial potentials makes them a natural first choice to apply our calculations to. Hence, we consider a general power-law inflaton potential of the form
\begin{equation}
    V(\phi) = \frac{\lambda_p}{p!}\frac{\phi^p}{M_\text{Pl}^{p-4}}\ , \label{potentialLargeField}
\end{equation}
where $\lambda_p$ is a dimensionless scalar self-coupling. 
The corresponding slow-roll parameters are
\begin{equation}
    \epsilon_V= \frac{1}{2}\left(\frac{pM_\text{Pl}}{\phi}\right)^2\ , \qquad \delta_V = p(p-1)\left(\frac{M_\text{Pl}}{\phi}\right)^2 \ ,\label{largeFieldSRparams}
\end{equation}
The requirement for slow-roll inflation is $\epsilon_H\simeq \epsilon_V<1$, which determines the end of inflation to be $\phi/M_\text{Pl}=p/\sqrt{2}$. We estimate the magnitude of the correlators at horizon exit for which 
\begin{equation}
    \left(\frac{\phi_*}{M_\text{p}}\right)^2 = 2pN\left(1+\frac{p}{4N}\right)\ . \label{phi_star}
\end{equation}
The number of $e$-foldings between horizon crossing and the end of inflation is given by the scale of inflation and the thermal history of the Universe after inflation. As indicated, we will simply assume that $N$ is in the interval $50-60$. 

The spectral index is given by
\begin{equation}
n_\text{s}-1=2\delta_H-4\epsilon_H\simeq 2\delta_V-6\epsilon_V=-\frac{p+2}{2N\left(1+\frac{p}{4N}\right)}.
\end{equation}
For $p=2$ and $N=57$ we get the observed value of $n_\text{s}\simeq 0.965$ \cite{Akrami:2018odb}. This determines $\epsilon_{H}\simeq 0.00875$ and $N_\text{t}\simeq 96$, $N_\text{s}\simeq 175$. 
The tensor correlator in Eq. \eqref{eq:tensorCorrMagGeneral} is then of the order
\begin{equation}
    C_{0}\simeq 4.8\frac{H^2}{M_\text{Pl}^2}\;.
\end{equation}
Similarly for the scalars, we use the relation of Eq. \eqref{eq:spectralIndexSRCombination} in Eq. \eqref{eq:scalarCorrMagGeneral}, to obtain
\begin{equation}
    D_{0} \simeq  2.2\frac{\epsilon_HH^2}{M_\text{Pl}^2}\ .
\end{equation}
For $p=4$, we need $N=85$ to get the correct spectral index. This is ruled out by observations, but for illustration, this then gives $\epsilon_H\simeq 0.012$, $N_\text{t}\simeq 259$, $N_\text{s}\simeq 507$ and
\begin{equation}
    C_{0}\simeq 13.1\frac{H^2}{M_\text{Pl}^2}\ ,\qquad D_{0} \simeq  6.4\frac{\epsilon_HH^2}{M_\text{Pl}^2}\ .
\end{equation}
In the mean-field equation (\ref{MeanFieldEqLeading}), we have
\begin{equation}
\begin{aligned}
    0 &= 3H\Dot{\phi}\left(1+C_0\right)
    +V_{,\phi}(1-C_0)+M_\text{Pl}^{2}V_{,\phi\phi\phi}\frac{D_0}{\epsilon_H}\ . 
\end{aligned}
\end{equation}
With $p=2$ then $V_{,\phi\phi\phi}=0$, and so scalar fluctuations give no corrections at leading order. The tensor modes do, although they are numerically very small.
For $p=4$ then $V_{,\phi\phi\phi}=\Lambda_4 \phi$, so both scalars and tensors contribute, and we may write
\begin{equation}
\frac{V_{,\phi}C_0}{M_\text{Pl}^{2}V_{,\phi\phi\phi}\frac{D_0}{\epsilon_H}}=\frac{2}{3}\frac{\phi^2_*}{M_{\rm Pl}^2}\frac{N_\text{t}}{N_\text{s}}\ ,
\label{eq:phicorrfinal}
\end{equation}
which in this case is $\sim \mathcal{O}(10^2)$. Hence, the tensor contributions dominate.

But of course, the overall magnitude of the corrections is controlled by the quantity $H^2/M^2_{\rm Pl}$, which is related to the overall amplitude of the CMB (scalar) spectrum and constrained to be strictly smaller than $(2.5\times 10^{-5})^2$ \cite{Akrami:2018odb}. Since $N_\text{s}$ is fixed by $n_{\rm s}-1$ and $N$, large corrections can only arise if $N_\text{t}$ would become large in a small-$\epsilon_H$ regime. In order for $C_0$ to be comparable to unity, then $N_\text{t}\simeq N$ must be of order $10^{10}$.
\begin{table}[ht]
\begin{center}
\begin{tabular}{c | c c c c} 
   & $\phi^2$ & $\phi^4$  & $V_0(1-\lambda_4\phi^4)|_{N=50}$ & $V_0(1-\lambda_4\phi^4)|_{N=60}$ \\ [0.5ex] 
 \hline
 $C_0$ [${H^2}/{M_\text{Pl}^{2}}$]& 4.8& 13.1& 2.8& 3.1\\
 \hline
 $D_0$ [${H^2}/{M_\text{Pl}^{2}}$]&$2.2\epsilon_H$ & $6.4\epsilon_H$& $1.7\epsilon_H$ & $2.6\epsilon_H$\\
 \hline
 $\Delta \rho$ [${H^4}/{M_\text{Pl}^{2}}$] & $-0.26$ & $-0.86$ & $-0.12$ & $-0.15$\\ 
 \hline
 $\Delta p$ [${H^4}/{M_\text{Pl}^{2}}$] & $-0.025$ & $-0.016$ & $-0.10$ & $-0.14$\\[1ex] 
\end{tabular}
\caption{Estimates of the tensor and scalar correlators ($C_0$ resp. $D_0$) as well as the Friedmann corrections in Eq. \eqref{eq:frcorrfinal} for two monomial potentials and the quartic hilltop potential of Eq. \eqref{smallFieldPotential}. For the quartic hilltop potential the parameter $\lambda_4$ is estimated given a selected value of the number of $e$-foldings $N$.}
\label{Table}
\end{center}
\end{table}

The corrections to the Friedmann equations become (taking $N_\text{t,s}$ to be constant)
\begin{equation}
    \begin{aligned}
    &\Delta\rho \equiv H^2\left[\left(-9\epsilon_H +3\delta_H\right)\frac{N_\text{s}}{8\pi^2} -\frac{19}{8}\epsilon_H \frac{N_\text{t}}{\pi^2}\right]\frac{H^2}{M_{\rm Pl}^2}\ ,\\
    &\Delta p \equiv H^2\left[\left(-9\epsilon_H +3\delta_H\right)\frac{N_\text{s}}{8\pi^2} +\frac{3}{8}\epsilon_H \frac{N_\text{t}}{\pi^2}\right]\frac{H^2}{M_{\rm Pl}^2}\ , 
\end{aligned}
\label{eq:frcorrfinal}
\end{equation}
for Eq. \eqref{FinalFriedmannTimeLead} and \eqref{FinalFriedmannSpaceLead} respectively. Estimates of these corrections are summarised in Table \ref{Table} for the potentials considered in this section. For both of the monomial potentials the corrections are of order $\Delta\rho\sim \mathcal{O}(10^{-1}{H^2}/{M_\text{Pl}^{2}})$ and $\Delta p\sim \mathcal{O}(10^{-2}{H^2}/{M_\text{Pl}^{2}})$, which are again  completely negligible. 

\subsection{Quartic hilltop}\label{sec:hilltop}

A general hilltop model \cite{Boubekeur:2005zm}  of the inflaton potential can be written
\begin{equation}
    V(\phi) = V_0\left[1-\lambda_p\left(\frac{\phi}{M_\text{Pl}}\right)^p+\dots\right]\ , \quad p\geq 2\ ,\label{smallFieldPotential}
\end{equation}
where $V_0$ is a constant energy density scale, $\lambda_p$ is some parameter, $p$ is often an integer, and the dots indicate that some higher power-law terms must kick in at larger values of $\phi$, for the potential to be bounded from below. We will take $p=4$.

The quantum contributions for this model can be estimated in much the same way as before. In fact, because $N_\text{s}$ is a function of the slow-roll parameters in the combination $2\delta_H-4\epsilon_H=n_\text{s}-1$, if we insist that the spectral index is the observed one, then $N_\text{s}$ is independent of the inflation model. 

Since the potential in Eq. \eqref{smallFieldPotential} is a constant plus a monomial, the expressions in Eqs. (\ref{eq:phicorrfinal}) and (\ref{eq:frcorrfinal}) will be left unchanged, and to find $N_\text{t}$ we only need to compute $\epsilon_H\simeq \epsilon_V$ at horizon crossing (since $\delta_H$ follows from $n_\text{s}-1$ and $\epsilon_H)$. 

With $p=4$, the potential slow-roll parameters are 
\begin{equation}
\epsilon_V(\phi) = \frac{8\lambda_4^2\left(\frac{\phi}{M_{\rm Pl}}\right)^6}{\left(1-\lambda_4\left(\frac{\phi}{M_{\rm Pl}}\right)^4\right)^2}\ ,\qquad \delta_V(\phi)=-\frac{12\lambda_4\left(\frac{\phi}{M_{\rm Pl}}\right)^2}{1-\lambda_4\left(\frac{\phi}{M_{\rm Pl}}\right)^4}\ ,
\end{equation}
and we define the end of inflation by $\epsilon_V(\phi_\text{e})\equiv 1$. From the definition of $N$ we then find
\begin{equation}
\left(\frac{\phi_*}{M_{\rm Pl}}\right)^2+\frac{1}{\lambda_4}\left(\frac{\phi_*}{M_{\rm Pl}}\right)^{-2}=
8N+\left(\frac{\phi_\text{e}}{M_{\rm Pl}}\right)^2+\frac{1}{\lambda_4}\left(\frac{\phi_\text{e}}{M_{\rm Pl}}\right)^{-2},
\end{equation}
from which we compute $\epsilon_V(\phi_*)$ and $\delta_V(\phi_*)$. For $N=50$ we find the values $\lambda_4=4.3\times 10^{-6}$, $(\phi_*)^2=200M_{\rm Pl}^2$, $N_\text{t}=55$ and $N_\text{s}=136$. With $N=60$ then we find $\lambda_4=3.2\times 10^{-5}$, $(\phi_*)^2=40M_{\rm Pl}^2$, $N_\text{t}=62$ and $N_\text{s}=205$. The fraction of Eq. (\ref{eq:phicorrfinal}) then becomes about $50$ for $N=50$ and $8$ for $N=60$. The Friedmann corrections of are of order $\Delta\rho,\Delta p\sim\mathcal{O}(10^{-1}{H^2}/{M_\text{Pl}^{2}})$ as listed in Table \ref{Table}.

\section{Conclusions and outlook}
\label{sec:conclusion}

We have computed the leading-order (in slow-roll and quantum fluctuations) corrections to the inflaton equation of motion and the Friedmann equations during inflation. We have included the fluctuations in the inflaton field, and both scalar and tensor fluctuations in the metric. 
Starting out including a non-minimal coupling to gravity $F(\phi)$, we solved the mode equation for the tensor modes, and were able to derive expressions for the scalar mode equations and the corrected Friedmann and field equations. We then proceeded in the limit $F(\phi)=1$ to explicitly compute the leading-order corrections to the evolution equations. 

We found that both the tensor and scalar correlators enter at leading order for a self-interacting inflaton, the scalars as an additive contribution, the tensors as multiplicative contributions. The tensors contribute also for a quadratic inflaton potential, where the scalar contribution decouples. While the tensor contribution is larger by one or two orders of magnitude, the overall size of the corrections is very small. 

The correlators are logarithmically IR-divergent, and we introduced an ad hoc IR-cutoff in order to estimate their magnitude. Infrared divergences in Minkowski space are well-known from finite-temperature and finite-density calculations. They are more severe in the former case due to phase-space effects (dimensional reduction), but in both cases they are unphysical in the sense that the real physical medium effects are screening them. This is resolved by resummation of classes of certain diagrams to all orders in perturbation theory, for example by using the resummation program of \cite{braaten} or the 2PI effective action formalism~\cite{2pi}. A mass can also be generated non-perturbatively in curved spacetime~\cite{Serreau:2011fu,Herranen:2013raa} that screens infrared divergences.\footnote{The mass is non-perturbative in the sense that it is not seen at any finite order in perturbation theory, but shows up after summing classes of diagrams to all orders.} For example, in de Sitter space, the generated mass for a classical massless scalar field is of order $H$. In Minkowski space, there are quantum fluctuations at all scales, however, in de Sitter space, the scale is set by $H$ and this is the scale of the important quantum fluctuations.

Since quantum corrections during inflation seems to anyway be suppressed by $H^2/M_{\rm Pl}^2$, computing them with high accuracy may seem futile. However, quantum and thermal corrections are known to be important for a number of phenomena in cosmology, with examples including infrared effects in de Sitter space \cite{Serreau:2011fu,Gautier:2013aoa,Gautier:2015pca}, corrections to the effective potential of the Standard Model \cite{Herranen:2014cua,Herranen:2015ima} and thermalisation \cite{Tranberg:2008ae}.

Although significantly more challenging than field theory in a classical FLRW background, exploring metric corrections to the field equations is part of that story. 

Much work remains to be done on resummation of the IR-divergences, including metric fluctuations and away from de Sitter space, as well as the further inclusion of non-minimal coupling to gravity and other more general theories of gravity. Also, although in this work we have insisted that $\phi$ is the inflaton, quantum corrections arise for spectator fields  as well (curvaton, other matter fields), due the background metric and its fluctuations. Although the Friedmann equations remain dominated by the inflaton, each subdominant spectator field acquires its own effective equation of motion. This opens up a number of new model-building opportunities to explore.

\appendix

\section{Scalar commutation relations}\label{Appendix:CommutationRelations}

For completeness, we list a number of commutation relations
in addition to those given in Eq. \eqref{eq:MainCorrelationRelationScalars}.
They are
\begin{equation}
\begin{aligned}
    [\pi_\Psi(\mathbf{x}),\pi_{\delta\phi}(\mathbf{x'})] &= i3a^2\left(\phi'+\frac{2(3\mathcal{H}'\phi'+a^2\mathcal{H}V_{,\phi})}{\nabla^2}\right)\delta^3(\mathbf{x}-\mathbf{x'})\ , \\
    [\Psi(\mathbf{x}),\pi_{\delta\phi}(\mathbf{x'})] &= -i\frac{3\mathcal{H}\phi'+a^2V_{,\phi}}{2M_\text{Pl}^2\nabla^2}\delta^3(\mathbf{x}-\mathbf{x'})\ , \\ [\delta\phi(\mathbf{x}),\pi_{\delta\phi}(\mathbf{x'})] &= i\left(1+\frac{3\phi'^2}{M_\text{Pl}^2\nabla^2}\right)\delta^3(\mathbf{x}-\mathbf{x'})\ , \\
    [\delta\phi(\mathbf{x}),\pi_{\Psi}(\mathbf{x'})] &= i\frac{6\mathcal{H}\phi'}{\nabla^2}\delta^3(\mathbf{x}-\mathbf{x'})\ , \\
    [\Psi(\mathbf{x}),\pi_{\Psi}(\mathbf{x'})] &= -i\frac{3\phi'^2}{2M_\text{Pl}^2\nabla^2}\delta^3(\mathbf{x}-\mathbf{x'})\ , \\
    [\Psi(\mathbf{x}),\delta\phi(\mathbf{x'})] &= i\frac{\phi'}{2M_\text{Pl}^2a^2\nabla^2}\delta^3(\mathbf{x}-\mathbf{x'})\ , \\
    [\delta\phi(\mathbf{x}),\delta\phi'(\mathbf{x'})] &= \frac{i}{a^2}\left(1+\frac{3\phi'^2}{2M_\text{Pl}^2\nabla^2}\right)\delta^3(\mathbf{x}-\mathbf{x'})\ ,
\end{aligned}\label{equaltimeCommutationRelations}
\end{equation}
where it is implicit that they apply at equal times. 

\section{Second-order equations with general non-minimal coupling}\label{Appendix:Feqs}

We present the Friedmann and mean-field equations  with a general non-minimal coupling to quadratic order in perturbations to illustrate the (rather non-trivial) dependence on $F$. A central ingredient to these equations is the perturbed inverse metric, which can be found from a Taylor expansion. To quadratic order in perturbations, we find 
\bqa\nonumber
    (g+\delta g)^{\mu\nu} &=& g^{\mu\nu}-\delta g^{\mu\nu}+\delta g^{\mu\lambda}
    \delta g_{\lambda}^{\,\,\,\nu} 
    \label{eq:perturbedInverseMetric}\\
    &=& \begin{pmatrix}
        1+2\Phi-4\Phi^2 && 0\\
        0 && \frac{1}{a^2}\left[\delta^{ij}+2(\Psi\delta^{ij}-E^{ij})+4\Psi^2\delta^{ij}
        +4E^{ik}E_{\,\,\,k}^{j}-8\Psi E^{ij}\right]
        \end{pmatrix}\;, \nonumber
        \\ &&
\eqa
where indices on the perturbed metric are raised as $\delta g^{\mu\nu}=-g^{\alpha\mu}g^{\nu\beta}\delta g_{\mu\nu}$.

Similarly, using $\det g = \exp(\Tr(\ln g))$ and Taylor expanding, the perturbed metric determinant is given by
\bqa\nonumber
    \sqrt{-(g+\delta g)} &=& \sqrt{-g}\left( 1+\tfrac{1}{2}\delta g^{\mu}{}_\mu
    +\tfrac{1}{8}\delta g^{\mu}{}_\mu \delta g^{\nu}{}_\nu-\tfrac{1}{4}\delta g^{\mu\nu}\delta g_{\nu\mu}\right)\\ \nonumber
    &=& a^3\big(1+\Phi-3\Psi+E^i{}_i-\tfrac{1}{2}\Phi^2-3\Phi\Psi+\Phi E^i{}_j\\
    &&
    +\tfrac{3}{2}\Psi^2-\Psi E^i{}_i+\tfrac{1}{2}
    E^i{}_iE^j{}_j-E^{ij}E_{ji}\big)\ , \label{eq:perturbedMetricDeterminant}
    \eqa
where $\delta g^{\mu}{}_{\mu}=g^{\mu\nu}\delta g_{\nu\mu}$. Eqs. (\ref{eq:perturbedInverseMetric}) and (\ref{eq:perturbedMetricDeterminant}) are then inserted into the expressions for the energy-momentum tensor in Eq. \eqref{eq:EMT} and mean-field equation \eqref{eq:EOMInflatonGeneral}, whose resulting expressions are truncated at quadratic order. 

With a general $F$, the perturbed Einstein field equations to quadratic order read
\begin{equation}
    (F+\delta F+\delta_2F)(G_{\mu\nu}+\delta G_{\mu\nu}+\delta_2 G_{\mu\nu})=\frac{1}{M_\text{Pl}^2}(\tilde{T}_{\mu\nu}+\delta\tilde{T}_{\mu\nu}+\delta_2\tilde{T}_{\mu\nu})\ , \label{eq:quadraticOrderPertEinsteinEq}
\end{equation}
from which the quantum-corrected equations take the form
\begin{equation}
    G_{\mu\nu} = \frac{1}{M_\text{Pl}^2}\frac{1}{F}\left(\tilde{T}_{\mu\nu}+\braket{\delta_2\tilde{T}_{\mu\nu}}\right)-\braket{\delta_2G_{\mu\nu}}-\left\langle\frac{\delta F}{F}\delta G_{\mu\nu}\right\rangle-\left\langle\frac{\delta_2F}{F}\right\rangle G_{\mu\nu} \ .  \label{eq:quantumCorrEinsteinEq}
\end{equation}
Specialising to the FLRW metric in Newtonian gauge, at linear order, the perturbations to the Einstein tensor read
\begin{equation}
\begin{aligned}
    {\delta G^0{}_0} &=  6H^2\Phi+6H\dot{\Psi}-\frac{2}{a^2}\nabla^2\Psi \ , \\
    {\delta G^i{}_j} &=  \left[6H^2\Phi+4\dot{H}\Phi+2H\dot{\Phi}+6H\dot{\Psi}+2\ddot{\Psi}+\frac{1}{a^2}\nabla^2(\Phi-\Psi)\right]\delta^i{}_j\\
    &\quad+\frac{1}{a^2}\partial_j\partial^i(\Psi-\Phi)+\ddot{E}^i{}_j+3H\dot{E}^i{}_j-\frac{1}{a^2}\nabla^2E^i{}_j\ ,
\end{aligned}
\end{equation}
and the quadratic-order contributions are given by 
\begin{equation}
\begin{aligned}
    {\delta_2G^0{}_0} &= \tfrac{1}{2} \dot{E}_{ij} \dot{E}^{ij} + 4HE^{ij} \dot{E}_{ij}  -12 H{}^2 \Phi{}^2 -12 H \Phi \dot{\Psi} + 12 H \Psi \dot{\Psi} -3 \dot{\Psi}^2 \\
    &\quad+\frac{1}{a^2}\big[- 8 \Psi \nabla^2\Psi - 3(\nabla\Psi)^2 + 2 E^{ij} \partial _ {j}\partial _ {i}\Psi - 2 E^{ij} \nabla^2E{}_{ij} + \partial _{j}E{}_ {ik} \partial ^{k}E^{ij} 
    - \tfrac{3}{2} \partial _{k}E{}_ {ij} \partial ^{k}E^{ij}\big]\ , \label{QuadraticEinsteinTensor00component}
\end{aligned}
\end{equation}
respectively
\begin{equation}
\begin{aligned}
{\delta_2G^i{}_j} &= -2 \dot{E}^{ik} \dot{E}_{jk} -2 E^{ik} \ddot{E}{}_{jk} + \tfrac{3}{2} \dot{E}_{kl} \dot{E}^{kl} \delta^i{}_j+ 2 E^{kl} \ddot{E}{}_ {kl} \delta^i{}_j-6H  E^{ik} \dot{E}_{jk} + 6H E^{kl} \dot{E}_ {kl} \delta^i{}_j \\
&\quad-2 \ddot{E}^{i}{}_ {j} \Phi -6 H \dot{E}^i{}_{j} \Phi -12 \delta^i{}_jH{}^2 \Phi{}^2 -8 \delta^i{}_j\dot{H} \Phi{}^2 - \dot{E}^i{}_ {j} \dot{\Phi}-8 \delta^i{}_jH \Phi \dot{\Phi}+ 2\ddot{E}^{i}{}_ {j} \Psi \\
&\quad+ 6 H  \dot{E}^i{}_{j}\Psi + \dot{E}^i{}_ {j} \dot{\Psi} + 6 H E^{i}{}_{j} \dot{\Psi} -12 \delta^i{}_jH \Phi \dot{\Psi} -2 \delta^i{}_j \dot{\Phi}\dot{\Psi} + 12 \delta^i{}_jH \Psi \dot{\Psi} + \delta^i{}_j \dot{\Psi}^2\\
&\quad+ 2 E^{i}{}_ {j} \ddot{\Psi}-4 \delta^i{}_j \Phi \ddot{\Psi} + 4 \delta^i{}_j \Psi \ddot{\Psi} + \frac{1}{a^2}\partial ^{i}E^{kl} \partial _{j}E{}_{kl} 
+ \frac{1}{a^2}\partial^{i}\Phi \partial _ {j}\Phi - \frac{1}{a^2} \partial ^{i}\Psi \partial _ {j}\Phi - \frac{1}{a^2}\partial ^{i}\Phi \partial _ {j}\Psi \\
&\quad+ \frac{3 }{a^2} \partial ^{i}\Psi \partial _ {j}\Psi + \frac{2}{a^2} E^{kl} \partial _ {j}\partial ^{i}E{}_{kl} + \frac{2}{a^2} \Phi \partial _ {j}\partial ^{i}\Phi 
- \frac{2}{a^2} \Psi \partial _ {j}\partial ^{i}\Phi + \frac{4}{a^2} \Psi  \partial _ {j}\partial ^{i}\Psi+  \frac{2}{a^2} E{}_{j}{}^{k} \partial _ {k}\partial ^{i}\Psi\\ 
&\quad+ \frac{2}{a^2} E^{ik} \partial _ {k}\partial _ {j}\Phi  - \frac{4}{a^2} \Psi \nabla^2E^{i}{}_{j} 
- \frac{2}{a^2} \delta^i{}_j \Phi \nabla^2\Phi + \frac{2}{a^2} \delta^i{}_j \Psi \nabla^2\Phi - \frac{2}{a^2} E^{i}{}_ {j} \nabla^2\Psi \\
&\quad- \frac{4}{a^2} \delta^i{}_j \Psi \nabla^2\Psi + \frac{1}{a^2}\partial ^{i}E{}_ {jk} \partial ^{k}\Phi + \frac{1}{a^2}\partial _{j}E^{i}{}_ {k} \partial ^{k}\Phi 
- \frac{1}{a^2}\partial _{k}E^{i}{}_ {j} \partial ^{k}\Phi - \frac{1}{a^2}\delta^i{}_j \partial _ {k}\Phi \partial ^{k}\Phi + \frac{1}{a^2}\partial ^{i}E{}_ {jk} \partial ^{k}\Psi \\
&\quad+ \frac{1}{a^2}\partial _{j}E^{i}{}_ {k} \partial ^{k}\Psi - \frac{3}{a^2} \partial _{k}E^{i}{}_ {j} \partial ^{k}\Psi - \frac{2 }{a^2} \delta^i{}_j \partial _ {k}\Psi \partial ^{k}\Psi - \frac{2}{a^2} E^{kl} \partial _ {l}\partial ^{i}E{}_{jk} - \frac{2}{a^2}  E^{kl} \partial _ {l}\partial_{j}E^{i}{}_{k}\\
&\quad+ \frac{2}{a^2} E^{kl} \partial _ {l}\partial _{k}E^{i}{}_{j} - \frac{2}{a^2} E^{kl} \delta^i{}_j \partial _ {l}\partial _ {k}\Phi + \frac{2}{a^2} E^{ik} \nabla^2E{}_{jk}- \frac{2}{a^2} \partial _{k}E{}_ {jl} \partial ^{l}E^{ik} \\
&\quad + \frac{2}{a^2} \partial _{l}E{}_ {jk} \partial ^{l}E^{ik} - \frac{2}{a^2}  E^{kl} \delta^i{}_j \nabla^2E{}_{kl}+ \frac{1}{a^2} \delta^i{}_j \partial _{l}E{}_ {km} \partial ^{m}E^{kl}- \frac{3}{2 a^2} \delta^i{}_j \partial _{m}E{}_ {kl} \partial ^{m}E^{kl}\ . \label{QuadraticEinsteinTensorIJcomponent}
\end{aligned}
\end{equation}
We keep in mind that since the tensor and scalar degrees of freedom propagate separately, the two-point correlators that mix these, e.g. $\braket{\delta\phi E_{ij}}$, vanish. 
Turning to the energy-momentum tensor, we determine the 00-component to
\begin{equation}
    \begin{aligned}
    \delta_2\tilde{T}^0{}_0 &= -\tfrac{1}{2}\dot{\delta\phi}^2-\tfrac{1}{2a^2}(\nabla\delta\phi)^2+2\dot{\phi}\Phi\dot{\delta\phi}-2\dot{\phi}^2\Phi^2-\tfrac{1}{2}V_{,\phi\phi}\delta\phi^2\\
    &\quad+3H\dot{\delta_2 F}-\frac{1}{a^2}\nabla^2\delta_2 F-3\dot{\delta F}(2H\Phi+\dot{\Psi})\\
    &\quad
    -\frac{1}{a^2}\left[\nabla(2\Phi-\Psi)\nabla\delta F+2\Psi\nabla^2\delta F-2\partial_i(E^{ij}\partial_j\delta F)\right]\\
    &\quad+2\dot{F}\big[ 6H\Phi^2+3(\Phi-\Psi)\dot{\Psi}-\dot{E}^i{}_jE^j{}_i \big]\ ,
\end{aligned}\label{QuadraticEMT00component}
\end{equation}
and for the $ij$-component, we get
\begin{equation}
    \begin{aligned}
    \delta_2\tilde{T}^i{}_j &= \frac{1}{a^2}\delta\phi^{,i}\delta\phi_{,j}+\big[\tfrac{1}{2}\dot{\delta\phi}^2-\tfrac{1}{2a^2}(\nabla\delta\phi)^2-2\dot{\phi}\Phi\dot{\delta\phi}+2\dot{\phi}^2\Phi^2-\tfrac{1}{2}V_{,\phi\phi}\delta\phi^2\big]\delta^i{}_j\\
    &\quad+ \frac{1}{a^2}\partial^i\partial_j\delta_2F+\big[\ddot{\delta_2 F}+2H\dot{\delta_2 F}-\frac{1}{a^2}\nabla^2\delta_2 F\big]\delta^i{}_j\\
    &\quad-2\ddot{\delta F}\Phi-\dot{\delta F}[(4H\Phi+\dot{\Phi}+\dot{\Psi})\delta^i{}_j+2\dot{E}^i{}_j]\\
    &\quad+\frac{1}{a^2}[\partial^i(-\Psi\delta^l{}_j+E^l{}_j)+\partial_j(-\Psi\delta^{li}+E^{li})-\partial^l(-\Psi\delta_j{}^i+E_j{}^i)]\partial_l\delta F\\
    &\quad+\frac{2}{a^2}(\Psi\delta^{ik}-E^{ik})\partial_k\partial_j\delta F-\frac{1}{a^2}\left[\nabla(\Phi-\Psi)\nabla \delta F+2\Psi\nabla^2\delta F-2\partial_k(E^{kl}\partial_l\delta F)\right]\delta^i{}_j\\
    &\quad+4\ddot{F}\Phi^2\delta^i{}_j+\dot{F}\big(8H\Phi^2+8\Phi\dot{\Psi}-2\Psi\dot{\Psi}+4\Phi\dot{\Phi}-2\dot{E}^k{}_lE^l{}_k\big)\delta^i{}_j\\
    &\quad+2\dot{F}\big[(\Phi-2\Psi)\dot{E}^i{}_j-2\dot{\Psi}E^i{}_j+2E^{ik}\dot{E}_{jk}\big]\ . 
\end{aligned}\label{QuadraticEMTIJcomponent}
\end{equation}
Finally, for the mean-field equation, we obtain 
\begin{equation}
\begin{aligned}
     0 &= \ddot{\phi}+3H\dot{\phi}+V_{,\phi}\\
     &\quad + \ddot{\phi}\big[\tfrac{3}{2}(\Phi+\Psi)^2+E^i{}_jE^j{}_i\big]+\dot{\phi}\big[ 3H\big(\tfrac{3}{2}(\Phi+\Psi)^2+E^i{}_jE^j{}_i\big)+3(\Phi+\Psi)(\dot{\Phi}+\dot{\Psi})+2\dot{E}^i{}_jE^j{}_i\big]\\
     &\quad-\ddot{\delta\phi}(\Phi+3\Psi)-\dot{\delta\phi}[3H(\Phi+3\Psi)+\dot{\Phi}+3\dot{\Psi}]-\frac{1}{a^2}\big(\nabla[(\Phi-\Psi)\nabla\delta\phi]-2\partial_i[E^{ij}\partial_j\delta\phi]\big)\\
     &\quad+\big(-\tfrac{1}{2}\Phi^2-3\Phi\Psi+\tfrac{3}{2}\Psi^2-E^i{}_jE^j{}_i\big)V_{,\phi}+(\Phi-3 \Psi)V_{,\phi\phi}\delta\phi+\tfrac{1}{2}V_{,\phi\phi\phi}\delta\phi^2    \\
     &\quad-\tfrac{1}{2}\left[\big(-\tfrac{1}{2}\Phi^2-3\Phi\Psi+\tfrac{3}{2}\Psi^2-E^i{}_jE^j{}_i\big)F_{,\phi}+(\Phi-3\Psi)\delta F_{,\phi}+\delta_2F_{,\phi}\right]R\\
     &\quad-\tfrac{1}{2}\left[(\Phi-3\Psi)F_{,\phi}+\delta F_{,\phi}\right]\delta R-\tfrac{1}{2}F_{,\phi}\delta_2R\ ,\label{MeanFieldEqBackgroundAndQuadraticTermsNonMinimal}
\end{aligned}
\end{equation}
where $\delta R$ is given in Eq. \eqref{deltaR} and the second-order perturbation to the Ricci scalar is given by 
\bqa\nonumber
    \delta_2 R &= &-3\dot{E}_{ij}\dot{E}^{ij}-4E_{ij}\ddot{E}^{ij}-16HE_{ij}\dot{E}^{ij}+\frac{1}{a^2}\left(4E_{ij}\nabla^2E^{ij}-2\partial_jE_{ik}\partial^kE^{ij}+3\partial_kE_{ij}\partial^kE^{ij}\right)\\
    && \nonumber
    +24(\dot{H}+2H^2)\Phi^2+24H\Phi\dot{\Phi}+48H(\Phi-\Psi)\dot{\Psi}+6\dot{\Phi}\dot{\Psi}+12(\Phi-\Psi)\ddot{\Psi}\\
    &&
    +\frac{2}{a^2}\left[2(\Phi-\Psi)\nabla^2\Phi+8\Psi\nabla^2\Psi+(\nabla\Phi)^2+3(\nabla\Psi)^2+\partial_i\Psi\partial^i\Phi\right]\; . 
    \label{Delta2R}
\eqa

\section{Correlator relations}\label{App:CorrRels}

The various quadratic-order terms in the Friedmann equations can be related to the two-point functions of the quantised fields via correlator relations. These relations can be obtained by differentiation and by use of the fields' equations of motion. 
The expectation values at quadratic order are defined to be symmetric. For the tensors we have that 
\begin{align}
    \braket{\dot{E}_{ij}E^{ij}} &\equiv \tfrac{1}{2}\braket{\dot{E}_{ij}E^{ij}+{E}_{ij}\dot{E}^{ij}} = \tfrac{1}{2}\partial_t\braket{{E}_{ij}E^{ij}}\ , \\
    \braket{\dot{E}_{ij}\dot{E}^{ij}} &= \tfrac{1}{2}\partial_t^2\braket{{E}_{ij}E^{ij}}-\tfrac{1}{2}\braket{\ddot{E}_{ij}{E}^{ij}+{E}_{ij}\ddot{E}^{ij}}\ , \\
    \braket{\ddot{E}_{ij}E^{ij}} &= -H(\tfrac{3}{2}+\epsilon_F)\partial_t\braket{{E}_{ij}E^{ij}}+\frac{1}{a^2}\braket{E_{ij}\nabla^2E^{ij}}\ , 
\end{align}
where the last relation is obtained from the tensors equation of motion \eqref{eq:TensorFieldEoM}. The scalar degrees of freedom are constrained via the field equations (\ref{LinearMixed}) and (\ref{LinearScalarAnisotropicStress}), that in the Einstein limit can be expressed in terms of the single field $\Psi$. The correlator relations of $\Psi$ are then obtained as
\begin{equation}
\begin{aligned}
    \braket{\dot{\Psi}\Psi} &= \tfrac{1}{2}\partial_tD_0 \ , \\
    \braket{\ddot{\Psi}\Psi} &= -\tfrac{1}{2}H(1+2\delta_H)\partial_tD_0-2H^2(\delta_H-\epsilon_H)D_0+\frac{1}{a^2}D_2\ , \\
    \braket{\dot{\Psi}^2} &= \tfrac{1}{2}\partial_t^2D_0-\braket{\ddot{\Psi}\Psi}\ , \\
    \braket{\ddot{\Psi}\dot{\Psi}} &= -H(1+2\delta_H)\braket{\Dot{\Psi}^2}-H^2(\delta_H-\epsilon_H)\partial_tD_0+\frac{1}{2a^2}\partial_tD_2\ , \\
    \braket{\dddot{\Psi}\Psi} &= -\tfrac{1}{2}\left[\partial_t[H(1+2\delta_H)]+2H^2(\delta_H-\epsilon_H)-H^2(1+2\delta_H)^2 \right]\partial_t D_0\\
    &\quad+\left[2H^3(1+2\delta_H)(\delta_H-\epsilon_H)-2\partial_t[H^2(\delta_H-\epsilon_H)]\right]D_0+\frac{1}{2a^2}\partial_tD_2-H(3+2\delta_H)\frac{1}{a^2}D_2\ , \\
    \braket{\ddot{\Psi}^2} &= H^2(1+2\delta_H)^2\braket{\Dot{\Psi}^2}+2H^3(1+2\delta_H)(\delta_H-\epsilon_H)\partial_tD_0+4H^4(\delta_H-\epsilon_H)^2D_0\\
    &\quad-H(1+2\delta_H)\frac{1}{a^2}\partial_tD_2-4H^2(\delta_H-\epsilon_H)\frac{1}{a^2}D_2+\frac{1}{a^4}D_4\ , \label{ScalarCorrelatorRelations}
\end{aligned}
\end{equation}
using the equation of motion for the scalar metric perturbation in Eq. \eqref{eq:MetricScalarEoM}. Again, all of the correlators should be understood to have equal time and space arguments. 

\section{Calculation of two-point correlators}\label{App:CalcCorrs}

From its mode decomposition in Eq. \eqref{eq:Eansatz}, the tensor correlator is given by
\begin{equation}
    C_0\equiv\braket{E_{ij}E^{ij}}=
    \int\frac{\diff^3q}{(2\pi)^{3/2}}\int\frac{\diff^3p}{(2\pi)^{3/2}}
    \sum_\lambda\sum_{\lambda'} e_{ij}^{(\lambda)}e^{(\lambda')*ij}\langle \hat{a}_{\lambda\mathbf{q}}\hat{a}^\dagger_{\lambda'\mathbf{p}}\rangle\tilde{h}_{\lambda\mathbf{q}}\tilde{h}_{\lambda'\mathbf{p}}\ , \label{EEcorrelator}
\end{equation}
where $\tilde{h}_{\lambda\mathbf{q}}$ is given in Eq. \eqref{tensorModeFunction}. We have already assumed that we are in the vacuum state, where the expectation value of the number operators $\hat{a}^\dagger_{\lambda\mathbf{q}} \hat{a}_{\lambda\mathbf{q}}$ vanish.
Then inserting
\begin{equation}
[\hat{a}_{\lambda\mathbf{q}},\hat{a}^\dagger_{\lambda\mathbf{q}}] =\delta_{\lambda\lambda'}\delta^3(\mathbf{q}-\mathbf{q'}),   
\end{equation}
we arrive at 
\begin{equation}
C_0\equiv\braket{E_{ij}E^{ij}}=
    \int\frac{\diff^3q}{(2\pi)^3}
    \sum_\lambda e_{ij}^{(\lambda)}e^{(\lambda)*ij} |\tilde{h}_{\lambda\mathbf{q}}|
    = \frac{4}{2\pi^2}\int \diff q\, q^2
     |\tilde{h}_{\lambda\mathbf{q}}|^2\; ,
     \label{c000}
\end{equation}
In the last line, we have summed over $i,j$
and $\lambda$ (giving a factor of 4) and used that the mode functions $\tilde{h}_{\lambda\mathbf{q}}$ only depend on the length of ${\bf q}$, and not on $\lambda$.

By introducing the new variable $x\equiv -q\eta$,  as $\eta<0$ inside the horizon, Eq. \eqref{EEcorrelator} can be written as
\begin{equation}
    C_0(\eta) = \frac{1}{4\pi}\frac{1}{M^2_{\rm Pl}a^2F}(-\eta)^{-2}\int_{\Lambda_{\text{IR}}}^{\Lambda_{\text{UV}}}\diff x x^2|H_\nu^{(2)}(-x)|^2\ , \label{EEcorrelatorInTermsOfX}
\end{equation}
where we have chosen a branch such that $|H_{\nu}^{(1)}(x)|^2=|H_{\nu}^{(2)}(-x)|^2$ for $x>0$.
We split the remaining radial integration interval into three  according to
\begin{equation}
    \int_{\Lambda_\text{IR}}^{\Lambda_\text{UV}}\diff x=\int_{\Lambda_\text{IR}}^{\kappa_\text{IR}}\diff x
    +\int_{\kappa_\text{IR}}^{\kappa_\text{UV}}\diff x
    +\int_{\kappa_\text{UV}}^{\Lambda_\text{UV}}\diff x\ , \label{integralsplit}
\end{equation}
with $\Lambda_\text{IR}\ll \kappa_\text{IR}\ll 1\ll \kappa_\text{UV}\ll \Lambda_\text{UV}$ so that we can use different approximations of the Hankel functions  $H_{\nu}^{(1)}(x)$ appropriate for each interval. In the low-momentum (IR) region,
we use the asymptotic expansion 
\begin{equation}
    H^{(1)}_{\nu,\text{IR}}(x)\equiv -\frac{i}{\pi}\left(\frac{2}{x}\right)^\nu\Gamma(\nu)+\dotso\ ,\label{IRexp}
\end{equation}
and in the intermediate-momentum region, we may let $\epsilon_{H,F}\rightarrow0$, so that the relevant Hankel function for the tensors reduces to
\begin{equation}
    H^{(1)}_{{\frac{3}{2}}}(x)= -\sqrt{\frac{2}{\pi x^3}}(i+x)e^{ix}\ . \label{eq:mediumexpTensors}
\end{equation}
In the large-momentum (UV) region we use the large-$|x|$ expansion of the Hankel function:
\begin{equation}
     H_{\nu,\text{UV}}^{(1)}(x)\equiv \frac{2}{\pi}\left[\frac{1}{x}-\frac{1-4\nu^2}{8x^3}\right]+\mathcal{O}(x^{-7/2})\; . \label{UVexp}
\end{equation}
Calculating the contributions from each momentum interval, the correlator in Eq. \eqref{EEcorrelator} becomes
\bqa\nonumber
    C_0\equiv\braket{E_{ij}E^{ij}}&=&\frac{1}{2\pi^2}\frac{H^2}{M_\text{Pl}^2F}\bigg[\left( \frac{1}{2(\epsilon_F+\epsilon_H)}+\log2+\psi(\tfrac{3}{2})\right)\left(-1+\Lambda_\text{IR}^{-2(\epsilon_H+\epsilon_F)} \right)\\
    &&+\tfrac{1}{2}\Lambda_\text{UV}^2+\log\Lambda_\text{UV}\bigg]+\mathcal{O}(\epsilon)    \;,
    \label{c00}
\eqa
where we have switched back to cosmic time and $\psi(x)$ denotes the digamma function. For the correlator involving two gradients, we have that 
\bqa\nonumber
        C_2 \equiv \braket{E_{ij}\nabla^2E^{ij}} &=& -4\int\frac{\diff^3q}{(2\pi)^{3}}q^2|\tilde{h}_{\lambda\mathbf{q}}|^2 \\
        &=& -\frac{1}{4\pi^2}\frac{a^2H^4}{M_\text{Pl}^2F}\left[\tfrac{1}{2}\Lambda_{\text{UV}}^4+\Lambda_{\text{UV}}^2-\tfrac{1}{2}\Lambda_\text{IR}^4-\Lambda_\text{IR}^2\right]\; ,
\label{c22}
\eqa
having performed the same procedure as for $C_0$. Note that keeping the non-minimal slow-roll parameter $\epsilon_F$ is straightforward for the tensors contributions.

For the scalars we work in the limit $F(\phi)=1$, so that the first correlator is given by
\begin{equation}
    D_0(\eta)\equiv\braket{\Psi^2} = \int\frac{\diff^3k}{(2\pi)^{3}}|f_\mathbf{k}|^2 = \frac{1}{16\pi}\frac{\epsilon_H\mathcal{H}^2}{M_\text{Pl}^2a^2}\int_{\Lambda_\text{IR}}^{\Lambda_\text{UV}}\diff x|H^{(2)}_{\upsilon}(-x)|^2 \;,
\end{equation}
and with the mode function given by Eq. \eqref{scalarModeFunction}. We then perform the same evaluation as for the tensors, with the difference being that the 
approximation to the Hankel function in the intermediate-momentum region reads
\begin{equation}
    H^{(1)}_{{\frac{1}{2}}}(x)= -\sqrt{\frac{2}{\pi x}}e^{ix} \ , \label{eq:mediumexpScalars}
\end{equation}
when $\epsilon_H,\delta_H\rightarrow0$. For the various loop integrals we then obtain
\begin{align}
\label{d00}
    D_0 &=\frac{\epsilon_HH^2}{ 8\pi^2 M_\text{Pl}^2}\left[\left( \frac{1}{2(2\epsilon_H-\delta_H)}+\log2+\psi(\tfrac{1}{2})\right)\left(-1+\Lambda_\text{IR}^{-2(2\epsilon_H-\delta_H)}\right)+\log\Lambda_\text{UV}\right]+\mathcal{O}(\epsilon)\ , \\
    D_2 &= -\frac{\epsilon_Ha^2{H}^4}{ 16\pi^2M_\text{Pl}^2}\left(\Lambda_\text{UV}^2-\Lambda_\text{IR}^2\right)+\mathcal{O}(\epsilon)\ , \\
    D_4 &= \frac{\epsilon_Ha^4{H}^6}{ 32\pi^2M_\text{Pl}^2}\left(\Lambda_\text{UV}^4-\Lambda_\text{IR}^4\right)+\mathcal{O}(\epsilon)\ .
    \label{d44}
\end{align}
The correlators in Eqs.~(\ref{c00}),~(\ref{c22}) and Eqs.~(\ref{d00})--(\ref{d44}) all contain powers and logarithms of the UV and IR cutoffs. Had we chosen to carry out the computation in dimensional regularisation, all powers would have immediately disappeared, leaving us only with the logarithmic divergences both in the UV and IR. The logarithmic UV divergences should be cancelled by counterterms, as in e.g. \cite{Markkanen:2013nwa}. The IR divergences, on the other hand, are physical in the sense that they indicate that a higher order effect (such as an infinite resummation) will generate a regulator (such as a gap) \cite{Sloth:2006az,Serreau:2011fu}.
With this in mind, we find it illuminating through the use of a simple cutoff regularisation to identify where divergence show up, but allow ourselves to assume that UV-divergences may be removed through a proper choice of counterterms. Simply discarding the UV-divergent terms amounts to a MS-like renormalisation condition.

\bibliographystyle{elsarticle-num} 

\end{document}